\documentclass[aps,twocolumn]{revtex4}
\usepackage{epsfig}
\newcommand{\be}{\begin{equation}}
\newcommand{\ee}{\end{equation}}
\newcommand{\bea}{\begin{eqnarray}}
\newcommand{\eea}{\end{eqnarray}}
\newcommand{\ba}{\begin{array}}
\newcommand{\ea}{\end{array}}

\begin{document}

\title{Simulations of lattice animals and trees}
\author{Hsiao-Ping Hsu, Walter Nadler, and Peter Grassberger}
\affiliation{John-von-Neumann Institute for Computing, Forschungszentrum
J\"ulich, D-52425 J\"ulich, Germany}
                                                                                
\date{\today}
\begin{abstract}
The scaling behaviour of randomly branched polymers in a good solvent 
is studied in two to nine dimensions, using as microscopic models lattice 
animals and lattice trees on simple hypercubic lattices. As a stochastic
sampling method we use a biased sequential sampling algorithm with
re-sampling, similar to the pruned-enriched Rosenbluth method (PERM) 
used extensively for linear polymers. Essentially we start simulating 
percolation clusters (either site or bond), re-weigh them according to 
the animal (tree) ensemble, and prune or branch the further growth 
according to a heuristic fitness function. In contrast to previous 
applications of PERM, this fitness function is {\it not} the weight with 
which the actual configuration would contribute to the partition sum,
but is closely related to it. We obtain high statistics of animals 
with up to several thousand sites in all dimension $2\leq d\leq 9$. In 
addition to the partition sum (number of different animals) we estimate 
gyration radii and numbers of perimeter sites. In all dimensions we 
verify the Parisi-Sourlas prediction, and we verify all exactly known
critical exponents in dimensions 2, 3, 4, and $\geq 8$. In addition, we present 
the hitherto most precise estimates for growth constants in $d\geq 3$. 
For clusters with one site attached to an attractive surface, we verify 
for $d\geq 3$ the superuniversality of the cross-over exponent $\phi$ 
at the adsorption transition predicted by Janssen and Lyssy, but not 
for $d=2$. There, we find $\phi=0.480(4)$ instead of the conjectured
$\phi=1/2$.
Finally, we discuss the collapse 
of animals and trees, arguing that our present version of the algorithm 
is also efficient for some of the models studied in this context, but 
showing that it is {\it not} very efficient for the `classical' model 
for collapsing animals.
\end{abstract}

\maketitle

\section{Introduction}

Lattice animals (or polyominoes, as they are sometimes called in mathematics
\cite{Golomb}) are just clusters of connected sites on a regular lattice.
Such clusters play an important role in many models of statistical 
physics, as e.g. percolation \cite{Stauffer92}, the Ising model 
(Fortuin-Kastleyn clusters, Swendsen-Wang algorithm \cite{Fortuin,Swendsen}),
and even lattice gauge theories~\cite{Drouffe}. The basic combinatorial problem
associated to them is to count the number $Z_N$ of different animals of 
$N$ sites. Two animals are considered as identical, if they differ just by
a translation (i.e., we deal with {\it fixed} animals in the notation of 
\cite{Jensen03}), but are considered as different, if a rotation or reflection
is needed to make them coincide. Thus there are e.g. $d$ animals of $N=2$
sites on a simple hypercubic lattice of dimension $d$, and $d(2d-1)$ animals
with $N=3$. 

The animal problem can be turned into a statistical problem by giving a 
statistical weight to every cluster. In contrast to percolation, where 
different shapes acquire different weights, all clusters with the same 
number $N$ of sites are given the same weight. This is similar to self 
avoiding walks (SAW). A SAW on a lattice is a connected cluster of 
$N$ sites with equal weight on all clusters, but with a restriction on 
its shape: each SAW has to be topologically linear, i.e. each site is
connected by bonds to at most 2 neighbours. No such constraint holds for
animals, thus animals are the natural model for randomly branched polymers
\cite{Lubensky}.

\begin{figure}
  \begin{center}
    \psfig{file=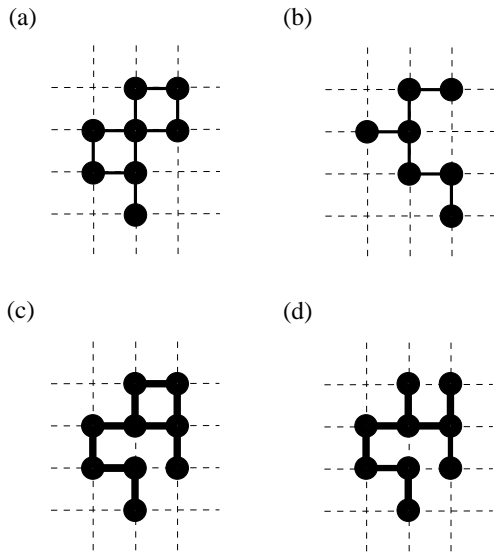,width=6.5cm,angle=0}
   \caption{(a): A site animal with 8 sites; (b) A site tree (``strongly embeddable tree");
    (c) A bond animal which is not a tree; (d) A bond tree (``weakly embeddable tree").}
\label{fig-examples}
\end{center}
\end{figure}

In addition to animals (or {\it site animals}, to be more precise) one can
also consider {\it bond animals} and {\it lattice trees}. A bond animal 
is a cluster where bonds can be established between neighbouring sites
(just as in SAWs), and connectivity is defined via these bonds: if there
is no path between any two sites consisting entirely of established bonds, 
these sites are considered as not connected, even if they are nearest 
neighbours. Different configurations of bonds are considered as different
clusters, and clusters with the same number of bonds (irrespective of their
number of sites) have the same weight \cite{Rensburg97}. {\it Weakly 
embeddable trees} are bond animals with tree topology, i.e. the set of 
weakly embeddable trees is a subset of bond animals, each with the same
statistical weight. {\it Strongly embeddable trees} are, in contrast, the
subset of site animals with tree-like structure. All these definitions are
illustrated in Fig.~\ref{fig-examples}.

Like many other statistical models, animals are characterized by scaling
laws in the limit of large $N$. It is believed that all the above 
statistics (site and bond animals, weakly and strongly embeddable trees)
are in the same universality class (same exponents, same scaling functions)
which is that of randomly branched polymers. The number of animals (i.e.
the microcanonical partition sum) should scale as \cite{Lubensky}
\be
    Z_N \sim \mu^N  N^{-\theta}\;,    \label{ZN}
\ee
and the gyration radius as
\be
       R_N \sim N^\nu \;,           \label{RN}
\ee
Here $\mu$ is the {\it growth constant} or inverse critical fugacity, 
and is not universal. In contrast, the Flory exponent $\nu$ and the 
entropic exponent $\theta$ should be universal.

In spite of the obvious similarity to the SAW and percolation 
problems, there are a number of features in which the animal 
problem is unusual:
\begin{itemize}

\item The upper critical dimension is $d=8$. There, $\nu=1/4$ and 
$\theta=5/2$ \cite{Adler}.

\item The model is not conformally invariant \cite{Miller93}, and thus the 
Flory exponent $\nu$ is not known exactly in $d=2$. 

\item Using supersymmetry, it has been argued by Parisi and Sourlas
~\cite{Parisi} that the animal problem in $d$ dimensions is related 
to the Yang-Lee problem (Ising model in an imaginary external field)
in $D=d-2$ dimensions. Based on this relationship (which is now 
proven rigorously \cite{Imbrie}, using a mapping onto the hypercubes 
problem at negative fugacity \cite{Baxter}) they argued that $\theta$ and $\nu$ 
should not be independent, but 
\be
         \theta=(d-2)\nu+1 \;.    \label{theta}
\ee
This implies in particular that $\theta=1$ in $d=2$. In addition, 
they showed that $\nu = 1/2$ in $d=3$.

\item Assuming universality so that scaling of the hard squares model at 
negative fugacity can be inferred from Baxter's solution of the hard 
hexagon model, and mapping the hard squares model onto 
lattice animals in 4 dimensions, Dhar \cite{Dhar} obtained 
$\theta=11/6$ for $d=4$. Thus one knows the exact values of $\nu$ for
$d=1,3,4$, and 8, but not for $d=2,5,6$, and 7.

\item In a series of papers, Janssen and Lyssy \cite{Janssen92,Janssen94,Janssen95}
studied animals attached to an adsorbing plane surface. For weak
adsorption (high temperature) the animals have basically the same 
structure as in the bulk, and the partition sum has the same scaling,
$Z'_N \sim \mu'^N N^{-\theta'}$ with $\mu'=\mu$ and $\theta' = 
\theta$ \cite{DeBell,DeBell92,footnoteDeBell}. For strong adsorption (low 
temperature) there is an adsorbed phase. Janssen and Lyssy argued
that the cross-over exponent between these two phases should be 
super-universal, $\phi = 1/2$ for all dimensions $d \geq 2$.

\end{itemize}

In the present paper we address all these points by means of a novel 
Monte Carlo algorithm which follows essentially the strategy used in 
the {\it pruned-enriched Rosenbluth method} (PERM) \cite{g97}. This 
is a recursively (depth first) implemented sequential sampling method with
importance sampling (bias) and re-sampling (``population control"). It seems
that PERM in the present implementation is much more efficient than previous
sampling methods for animals and trees. Indeed we shall present numerous
new estimates for critical exponents and growth constants which had 
previously been measured only with much larger error bars or not at all.

All this holds for athermal animals and trees, i.e. when there are no 
attractive forces between monomers. When such forces become strong, a number
of different collapse phase transitions are claimed to occur, depending 
on the detail of the model 
\cite{Derrida-H,Dickman,Lam87,Lam88,Madras90,Flesia92,Flesia94,Seno94,Henkel96,Stratychuk95,Madras97,Rensburg99,Rensburg00}. 
The simplest one of these involves site animals and a simple contact energy 
for each occupied nearest neighbour pair \cite{Derrida-H,Dickman,Lam87,Lam88} and 
is undisputed. But another transition, between two collapsed phases with 
different densities of bonds \cite{Madras90,Flesia92,Flesia94,Madras97,Rensburg00}, 
is still controversial \cite{Seno94,Henkel96,Stratychuk95}. At present 
all versions of our algorithm become inefficient for the first model, when 
the collapse point is approached. This is a bit disappointing in view of 
the fact that PERM for linear polymers is dramatically {\it more} efficient 
at the coil-globule transition than for athermal SAWs \cite{g97}. 
Obviously this leaves much room for further improvements. On the other 
hand, our method should work well for the other transition in large parts 
of the phase diagram.

Details of the algorithm for site 
animals will be given in Sec.~2. Detailed studies of site animals in the
bulk and in contact with a wall will be presented in Secs.~3 and 4. Bond
animals and trees will be discussed in Sec.~5. Finally, in Sec.~6 we will
study animal (and tree) collapse due to attractive forces between monomers.
The paper ends with conclusions in Sec.~7.

\section{Numerical Methods}

\subsection{Previous Methods}

\subsubsection{$\epsilon$-Expansions}

Field theoretic $\epsilon$-expansions (where $\epsilon$ is the distance 
from the critical dimension) were applied already very early to animals 
\cite{Lubensky} and to the Yang-Lee problem \cite{Alcantara}. When the 
relationship between both problems was established, the latter 
gave the most precise predictions for critical animal behaviour 
in high dimensions ($d \geq 5$).

\subsubsection{Exact Enumeration}

Exact enumeration of animals and trees is surprisingly non-trivial
\cite{Martin,Redner79,Redelmeier,Mertens}. Nevertheless, very extensive 
enumerations have recently been performed by 
I. Jensen~\cite{Jensen00,Jensen01,Jensen03} for site animals
and site trees in $d=2$. At present they give the best numerical 
verification of the prediction $\theta = 1$, and they give the most 
precise estimates for the Flory exponent ($\nu = 0.64115\pm 0.00005$) 
and for the growth constants:
$\mu=4.0625696\pm 0.0000005$ for animals \cite{Jensen03}, 
and $3.795254\pm 0.000008$ for trees. These 
values are more precise than old estimates obtained by finite size
scaling using strip geometry \cite{Derrida-S}. There are also 
enumerations of various animals and trees in higher dimensions 
\cite{Peters,Sykes,Gaunt82,Whittington,Lam,Adler,Madras90,Edwards,Foster}, 
but they are much less complete and in general they do not at present 
give the best estimates for critical parameters.

\subsubsection{Markov Chain Monte Carlo Methods}

The latter are obtained nowadays by Markov chain Monte Carlo (MCMC)
methods. Such algorithms have been used for animals since at least 20 
years \cite{Stauffer,Seitz,Dickman,Glaus}. At present, the most efficient 
MCMC algorithm for lattice trees is a version of the pivot algorithm 
\cite{Rensburg92,Rensburg97,You98,You01,Rensburg03}.
These simulations showed that $\nu=1/2$ in $d=3$, as predicted by 
\cite{Parisi}. Simulations of animals attached to an attractive wall 
verified that indeed $\phi=1/2$ in $d=2$ \cite{You01} (as also
verified with transfer matrix and similar methods \cite{Queiroz,Vujic}), 
although simulations in $d=3$ gave $\phi\approx 0.714$ \cite{Lam-Binder}, 
in gross violation of the Janssen-Lyssy prediction.

When applied to SAWs, the
pivot algorithm works by choosing a pivot point and proposing a rotation
of the shorter arm around the pivot, and accepting it when this leads to
no violation of self avoidance \cite{Madras}. When adapted to trees, 
one again chooses a random pivot point, but now the entire branch hinging
on this pivot is cut and glued somewhere else. Again this move is accepted
only if this leads to no violation of self avoidance and if it would not
lead to wrong cluster topology.

This method also allows to estimate growth constants, if it is used 
together with the atmosphere method \cite{Rensburg03}. In the latter, 
it is
counted how many possible ways there exist to grow the cluster by one 
further step, giving in this way an estimate of $Z_{N+1}/Z_N$. Basically 
the same method had been used in \cite{Grassberger03} to obtain very precise
estimates for the critical percolation thresholds in high dimensional 
lattices.

\subsubsection{Cluster Growth (`Sequential Sampling', `Static') Methods}

The first stochastic growth algorithm for trees seems to have been
devised by Redner~\cite{Redner}. Similar methods were then used by 
Meirovitch~\cite{Meirovitch} and Lam~\cite{Lam} for animals. But already 
Leath seems to have realized that his well known algorithm for growing 
percolation clusters \cite{Leath} could be used also for the study of animals, 
simply by reweighing the clusters. Recently this was taken up systematically 
in \cite{Care}. 

In the following we shall discuss the latter in some detail, and 
we shall restrict our discussion to site animals. The authors of 
\cite{Care} basically use a standard growth algorithm for percolation
clusters \cite{Leath,Grassberger83,Swendsen2}, and then reweigh the 
cluster so that they obtain the correct weight for the animal ensemble.
In a percolation cluster growth algorithm for site percolation, one 
starts with a single seed site and writes it into an otherwise empty 
list of `growth sites'. Then one recursively picks one item in the list 
of growth sites, removes it from the list and adds it with probability 
$p$ to the cluster,
and adds all its wettable neighbours to the list. This is repeated  
until either the cluster size exceeds some fixed limit (in which case
the cluster is discarded), or until the list is empty. A cluster with 
$N$ sites, $b$ boundary sites, and with fixed shape is obviously obtained
with probability 
\be
   P_{Nb} = p^N(1-p)^b,                     \label{Pperc}
\ee
i.e. with the correct probability so that
any unweighed average is just an average over the percolation ensemble.
Repeating this many times, the animal partition sum is then
\be
   Z_N = \langle 1/P_{Nb}\rangle = p^{-N} \langle (1-p)^{-b}\rangle\;.
\ee

The authors of \cite{Care} called their method a Rosenbluth method, 
in view of the obvious analogy with the Rosenbluth-Rosenbluth method
\cite{Rosenbluth55} for SAWs. In the latter one also samples from a biased
ensemble and then reweighs each configuration with the inverse of the 
sampling probability to obtain the correct partition sum. 

\subsection{PERM}

Like the 
original Rosenbluth-Rosenbluth method for SAWs, the method of \cite{Care}
has the disadvantage that the weights have a very wide distribution for 
large $N$. Thus even a very large sample will finally, when $N$ gets too
big, be dominated by a single configuration, and the method becomes 
inefficient even though it is easy to generate huge samples.
 
PERM (or any other strategy with resampling) tries to avoid this by 
trimming 
the width of the distribution of weights, by pruning configurations with
very low weight and making clones of high weight configurations which 
then share the weight among themselves. In many situations this has 
proven to be extremely efficient \cite{permreview,star1,star2}. But we 
cannot yet apply it to animals, since we 
have to be able to estimate the weight of a cluster {\it while it is 
still growing}, and up to now we have only discussed the relationship
between animals and percolation clusters after they had stopped growing.

In the following we shall again discuss only site animals, bond animals 
and trees being discussed in Sec.~5.

To obtain the relationship between still growing percolation clusters 
and animals, let us consider a cluster with $N$ sites, $g$ growth sites,
and $b$ sites which definitely belong to the boundary. At each of the 
growth sites the cluster can grow further, or it can stop growing with
probability $1-p$. Thus this cluster will contribute with weight $(1-p)^g$
to the sample of percolation clusters with $N$ sites and $b+g$ boundary
sites. Its contribution to the animal ensemble is smaller by a factor 
$[p^N(1-p)^{b+g}]^{-1}$, and we have thus
\be
   Z_N = p^{-N} \langle (1-p)^{-b}\rangle\;.     \label{reweight}
\ee
This is exactly the same formula as above, but now the average is taken 
over all growing clusters, while before we had averaged over clusters 
which had stopped growing.

Before we can implement these ideas, we have to discuss two problems:\\
(a) How are the clusters to be grown precisely?\\
(b) On what basis should we decide when to prune and when to clone a 
cluster?\\
As we shall see, both questions are not trivial.

\begin{figure}
  \vglue -5.mm
  \begin{center}
   \psfig{file=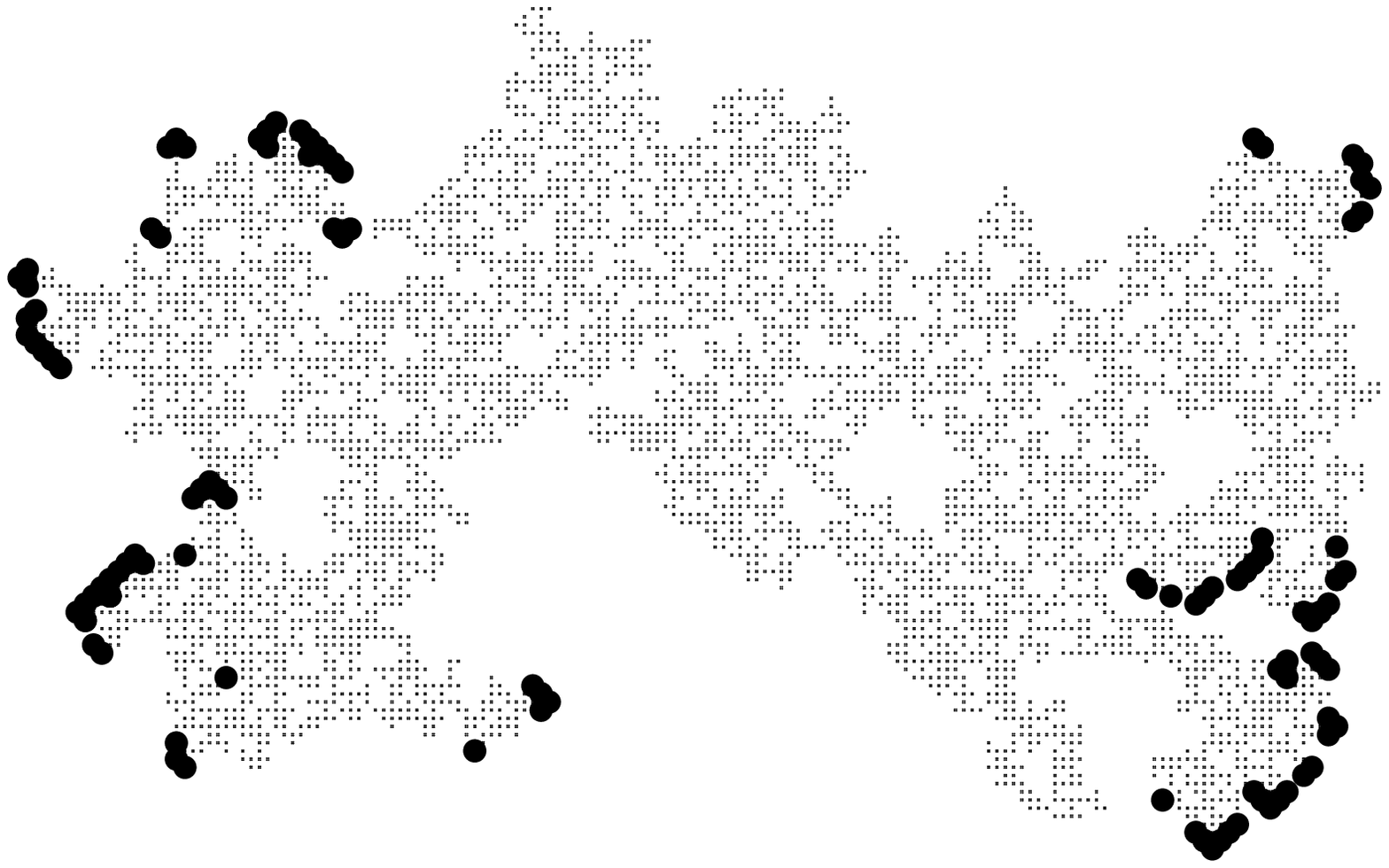,width=5.8cm,angle=270}
   \psfig{file=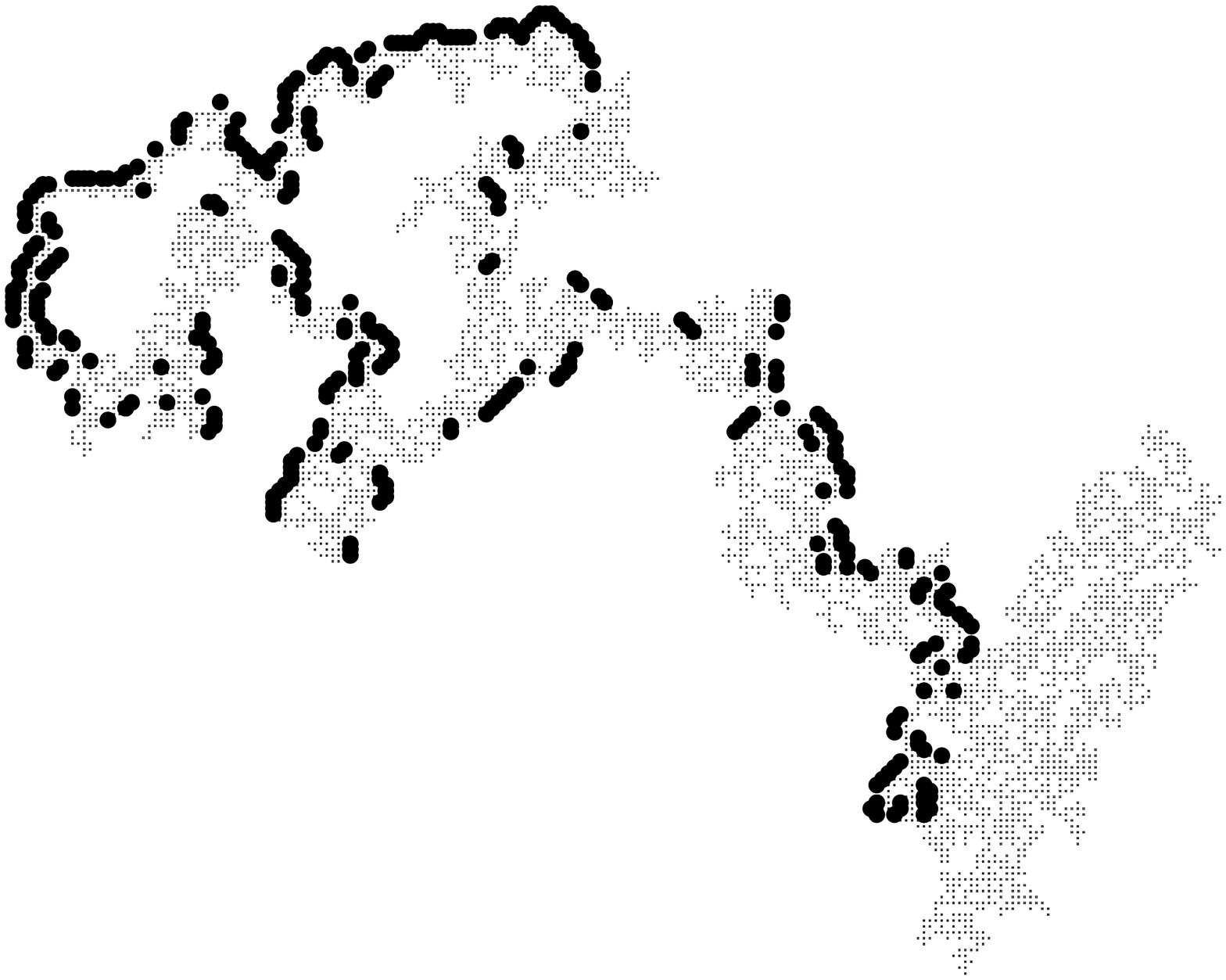,width=5.8cm,angle=270}
   \caption{Growing clusters generated breadth first (top)
   and depth first (bottom). In both cases we used $p=p_c=0.5927$, and in 
   both cases $N=4000$. Occupied sites are depicted by small points, 
   growth sites by heavy dots. Both figures are plotted with the same 
   scale.}
\label{leath}
\end{center}
\end{figure}

(a) Percolation cluster growth algorithms can be depth first or breadth
first. In the former, growth sites are written into a first-in last-out
list (a {\it stack}). A typical code for this is given in \cite{Swendsen2}.
For a breadth first implementation we use, instead, a first-in first-out
list (`fifo-list' or {\it queue}). Two 2-$d$ clusters growing according 
to these two schemes, with $p=p_c=0.5927$, are shown in 
Fig.~\ref{leath}. Both have $N=4000$. But the cluster grown using a stack
has a completely different shape and has $\approx 3$ times as many growth 
sites as the one grown with a queue! Most of these growth sites are nearly 
dead (their descendents will die after a few generations), but this is not 
realized because they are never tested. Since also
the fluctuations in the number of growth sites are much bigger in a depth
first implementation, the weights in Eq.(\ref{reweight}) will also 
fluctuate much more, and we expect much worse behaviour. This is indeed
what we found numerically: Results obtained when using a stack for the 
growth sites were dramatically much worse than results obtained with a 
queue. 

Notice that this is independent of the way how pruning and cloning is 
done. Indeed we implemented this ``population control" recursively as a
depth first strategy, as was done for all previous applications of PERM.

In addition, there are also some minor ambiguities in percolation cluster
growth algorithms, such as the order in which one searches the neighbours
of a growth site and writes them into the list. In 2 dimensions one can 
e.g. use the preferences east-south-west-north, or east-west-north-south,
or a different random sequence at every point. We found no big differences
in efficiency.

(b) In most previous applications of PERM, the best strategy was to base
the decision whether to prune or branch directly on the weight with which
the configuration contributes to the partition sum \cite{footnote0}.
This would mean in 
the present case that we clone, if $W_n \equiv p^{-n}(1-p)^{-b} > c_+ 
\hat{Z_n}$ where $c_+$ is a constant of the order $1\ldots 10$ and 
$\hat{Z_n}$ is the current estimate of $Z_n$. Similarly, a cluster would
be killed (with probability 1/2 \cite{g97}), if $W_n < c_- \hat{Z_n}$ 
with $c_-$ slightly smaller than 1.

In the present case this would not be optimal, since it would mean that
mostly clusters with few growth sites are preferred (they tend to have 
larger values of $b$, for the same $n$), and these clusters 
would die soon and would contribute little to the growth of much larger
clusters. Thus we defined a {\it fitness function}
\be 
   f_n = W_n /(1-p)^{\alpha g} = p^{-n}(1-p)^{-b-\alpha g}
\ee
with a parameter $\alpha$ to be determined empirically, and used 
\be
   f_n > c_+ \langle f_n \rangle, \qquad f_n< c_-  \langle f_n \rangle
\ee
as criteria for cloning and pruning. We checked in quite extensive
simulations that best results were obtained with $\alpha=1$ (except
when $N$ is small), and in the following we shall use only this
choice.

Finally we have to discuss the optimal values of $p$. It is clear 
that we should not use $p>p_c$, where $p_c$ is the critical percolation
threshold. Since minimal reweighing is needed for small $p$ (subcritical
percolation is in the animal universality class), one might 
expect $p\ll p_c$ to be optimal. This is indeed true for small values
of $N$ (which we are not primarily interested in), but not for large 
$N$. For the latter it is more important that clusters grown with 
$p\ll p_c$ have to be cloned excessively, since they otherwise would die 
rapidly in view of their few growth sites. 

\begin{figure}
  \begin{center}
   \psfig{file=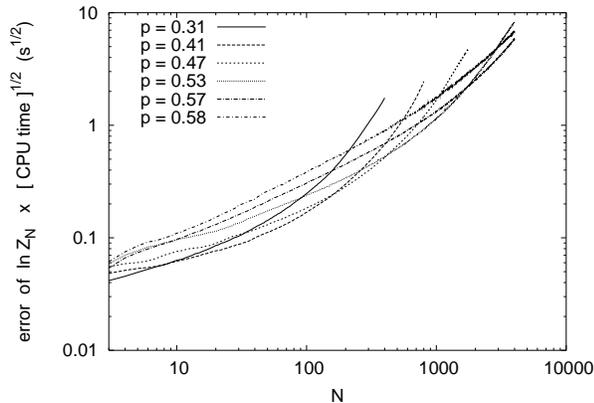,width=5.5cm,angle=270}
   \caption{Statistical errors of $\ln Z$ for lattice animals in $d=2$ 
   for various values of $p$. To make the different runs comparable, 
   errors are multiplied by the square root of the CPU time measured in 
   seconds. In 2 dimensions $p_c = 0.5927$.}
\label{err2d}
\end{center}
\end{figure}

\begin{figure}
  \begin{center}
    \psfig{file=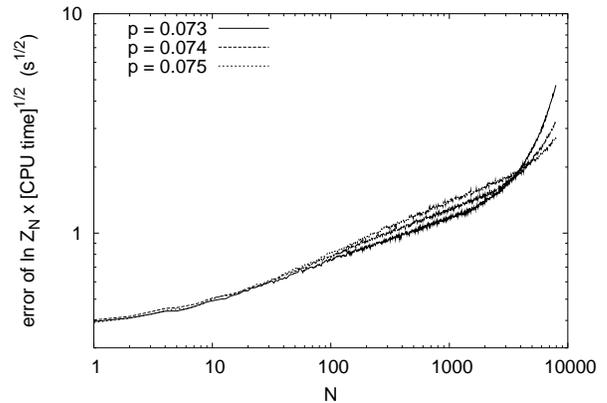,width=5.5cm,angle=270}
   \caption{Same as Fig.~\ref{err2d}, but for $d=8$ where $p_c = 0.0752$. The 
    straightest curve corresponds to the largest $p$.}
\label{err8d}
\end{center}
\end{figure}

To decide this problem empirically, we show in Fig.~\ref{err2d} the errors 
of the estimated free energies $F_N = -\ln Z_N$ for $d=2$. More precisely, 
we show there one standard deviation multiplied by the square root of the 
CPU time (measured in seconds), for different values of $p$. Each 
simulation was done on a Pentium with 3 GHz using the gcc compiler under 
Linux, and each simulation was done for $N_{\rm max}=4000$ (although we 
plotted some curves only up to smaller $N$, omitting data which might not 
have been converged). We see clearly that small values of $p$ are good 
only for small $N$. As $N$ increases, the best results were obtained for 
$p\to p_c$. The same behaviour was observed also in all other dimensions, 
and also for animals on the bcc and fcc lattices in 3 dimensions (data
not shown). As an example we 
show in Fig.~\ref{err8d} the analogous results for $d=8$. There we used a 
64 bit machine (a 600 MHz Compaq ALPHA), because this simplified hashing
(for large $d$ we used hashing as described e.g. in \cite{Grassberger03}).

\begin{figure}
  \begin{center}
    \psfig{file=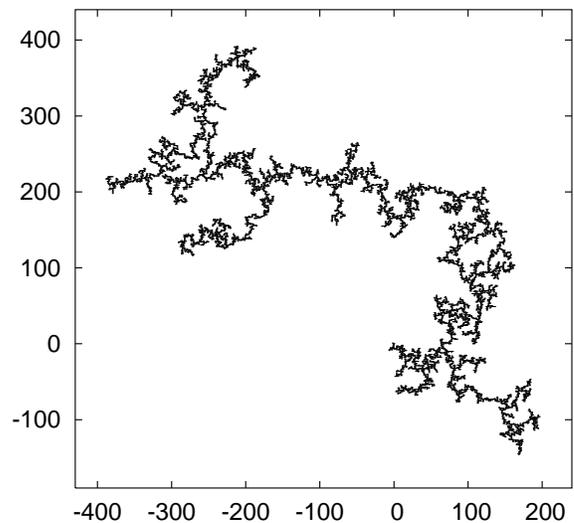,width=7.4cm,angle=270}
   \caption{A typical lattice animal with $12000$ sites on the square
    lattice.}
\label{animal2d}
\end{center}
\end{figure}

\begin{figure}
  \begin{center}
    \psfig{file=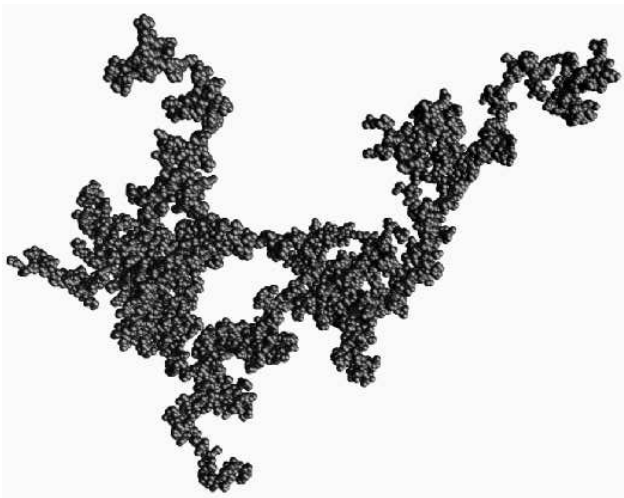,width=6.6cm,angle=270}
   \caption{A 3-$d$ lattice animal with $16000$ sites on the bcc lattice.}
\label{animal3d}
\end{center}
\end{figure}

Notice that the errors shown in Fig.~\ref{err8d} are much smaller than 
those in Fig.~\ref{err2d}, although the machine was slower and the animals
were twice as large ($N_{\rm max}=8000$). Indeed, the errors decreased
monotonically with $d$, being largest for $d=2$. Using $p$ slightly smaller
than $p_c$ we can
obtain easily very high statistics samples of animals with several thousand
sites for dimensions $\geq 2$. A typical 2-$d$ animal
with 12000 sites is shown in Fig.~\ref{animal2d}, and a 3-$d$ animal on the
bcc lattice with 16000 sites is shown in Fig.~\ref{animal3d}.

\begin{figure}
  \begin{center}
   \psfig{file=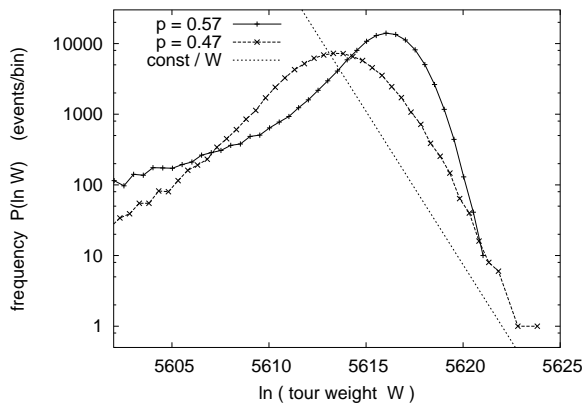,width=5.5cm,angle=270}
   \caption{Log-log plot of distributions of tour weights of $2d$ animals
    with $N=4000$, for $p=0.57$ and for $p=0.47$, together with a straight line
    representing the function $y=const/W$.}
\label{tourw2d}
\end{center}
\end{figure}

To check the reliability of our error bars we looked at distributions
of {\it tour weights} as described in \cite{Grass99}. A tour is the set
of all configurations generated by cloning from one common start and 
therefore possibly being strongly correlated. If the distribution $P$ of
tour weights $W$ is very broad, we are back to the problem of the Rosenbluth
method that averages might be dominated by a single tour. To check for this,
we plot $P(\ln W)$ against $\ln W$, and compare its right hand tail to
the function $1/W$. If the tail decays much faster, we are presumably on
the safe side, because then the product $W P(\ln W)$ has its maximum
where the distribution is well sampled. If not, then the results can still
be correct, but we have no guarantee for it.

In Fig.~\ref{tourw2d}, we show these tour weight distributions for 
two-dimensional animals with 4000 sites, for $p=0.57$ and for $p=0.47$ 
\cite{footnote1}. We see that the 
simulation with $p=0.57$ is on the safe side, but not that for $p=0.47$.
Similar plots for other simulations described in this paper showed that
all results reported below are converged and reliable.

Error bars quoted in the following on raw data (partition sums, 
gyration radii, and average numbers of perimeter sites or bonds) are 
straightforwardy obtained single standard deviations. Their estimate 
is easy since clusters generated in different tours are independent, 
and therefore errors can be obtained from the fluctuations of the 
contributions of entire tours (notice that clusters within one tour 
are {\it not} independent, and estimating errors from their individual
values would be wrong).

On the other hand, errors on critical exponents and on growth constants
are obtained by extrapolation. This is an ill-posed problem, and therefore
any error obtained this way is to some degree subjective. All such 
errors quoted in the following are based on plotting the data in different 
ways, plotting effective exponents against different powers of $1/N$, 
trying different ansatzes for higher order correction to scaling terms, 
etc. They are {\it not} based on simply making least square fits over 
fixed intervals of $N$, as this could lead to very large underestimations
of corrections to scaling. All quoted numbers are such that we believe,
to the best of our knowledge, that the true value is most likely within
one quoted error bar.

The total CPU time spent on the simulations reported in this paper is 
$\approx 25,000$ hours on fast PCs and work stations.

\section{Site Animals in 2 to 9 Dimensions}

\subsection{$d=2$}

\begin{figure}
  \begin{center}
  \psfig{file=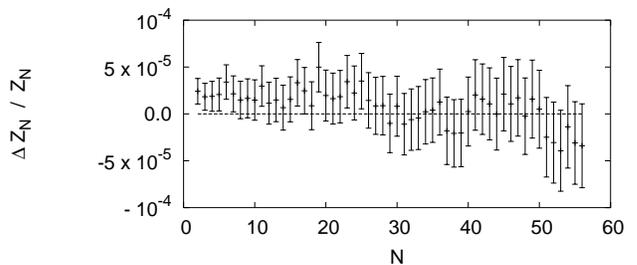,width=3.6cm,angle=270}
   \caption{Deviations of the logarithm of the number of 2-$d$ site animals
    from exact enumerations of I. Jensen~\cite{Jensen01}.}
\label{dz2d}
\end{center}
\end{figure}

\begin{figure}
  \begin{center}
    \psfig{file=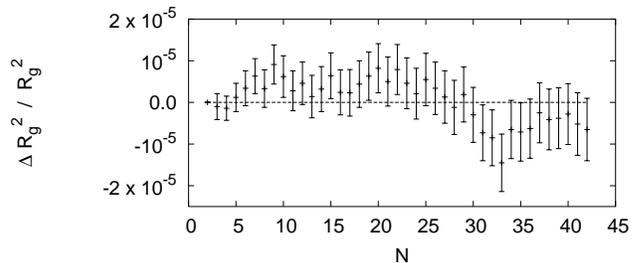,width=3.6cm,angle=270}
   \caption{Relative deviations of the squared radii of gyration from 
    exact values of \cite{Jensen01}.}
\label{dr2d}
\end{center}
\end{figure}

Before we report our final results, we show one more test where we compare our 
raw data for $d=2$ with the exact enumerations of \cite{Jensen03}. 
In Fig.~\ref{dz2d} we show the true relative errors of our estimates of 
the partition sum. Although there is some systematic trend visible, this is 
still within two standard deviations and thus not significant (notice that 
our values for different $N$ are not independent). Relative errors of the 
squared gyration radii are shown in Fig.~\ref{dr2d}. These data show 
that our estimates are basically correct, including the error bars.

\begin{figure}
  \begin{center}
    \psfig{file=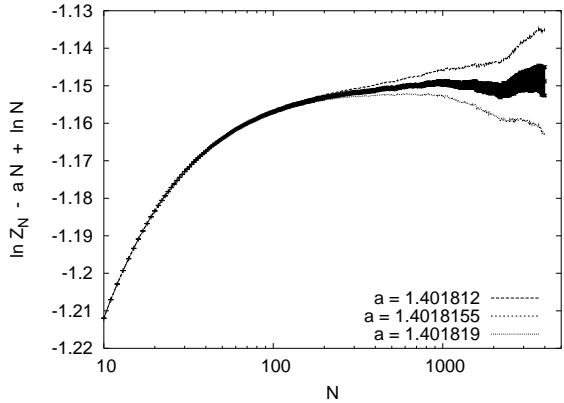,width=5.5cm,angle=270}
   \caption{Plot of $\ln Z_N-aN + \theta \ln N$ against $\ln N$ for $d=2$, 
    with $a=1.401812, 1.4018155$, and 1.401819 (top to bottom). Error
    bars are plotted only for the central curve.}
\label{zn2d}
\end{center}
\end{figure}

\begin{figure}
  \begin{center}
    \psfig{file=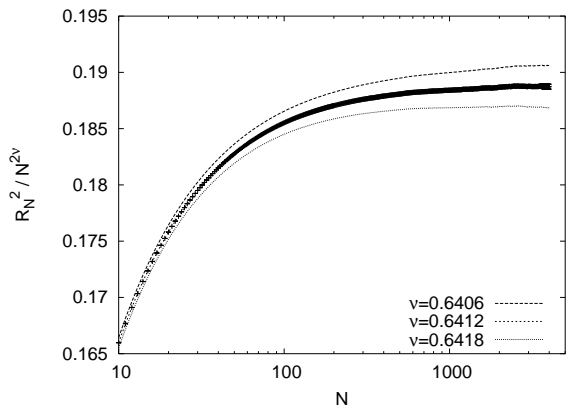,width=5.5cm,angle=270}
   \caption{Plot of $R_N^2/N^{2\nu}$ against $\ln N$ for $\nu=0.6406, 0.6412$, 
    and 0.6418. Again, error bars are plotted only for the central curve.}
\label{rn2d}
\end{center}
\end{figure}

Plotting directly our values of $Z_N$ would not be very informative, neither 
would be a plot of $\ln Z_N - a N$, where $a=\ln \mu$. Both ways of plotting
would hide any statistical errors. A more meaningful way of plotting our full 
data for $Z_N$ is used in Fig.~\ref{zn2d}, where we plot $\ln Z_N-a N + \ln N$ 
against $N$ for three values of $a$. Error bars are shown only 
for the central curve, although all three curves have of course the same errors.
In view of Eq.~(\ref{ZN}), and accepting the prediction that $\theta=1$, we 
would expect a curve which becomes horizontal for large $N$. This is indeed 
seen for the central curve, but the obvious corrections to scaling make a 
precise estimate of $\mu$ difficult. The same is true for the gyration radii.
In Fig.~\ref{rn2d} we show $R_N^2/N^{2\nu}$ against $N$ for three candidate 
values of $\nu$. Again strong corrections to scaling are seen. 

\begin{figure}
  \begin{center}
   \psfig{file=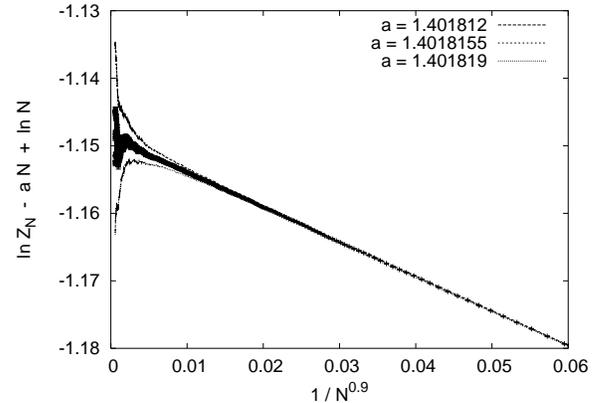,width=5.5cm,angle=270}
   \caption{Same data as in Fig.~\ref{zn2d}, but plotted against $1/N^{0.9}$.}
\label{zn2da}
\end{center}
\end{figure}
                                                                               
\begin{figure}
  \begin{center}
   \psfig{file=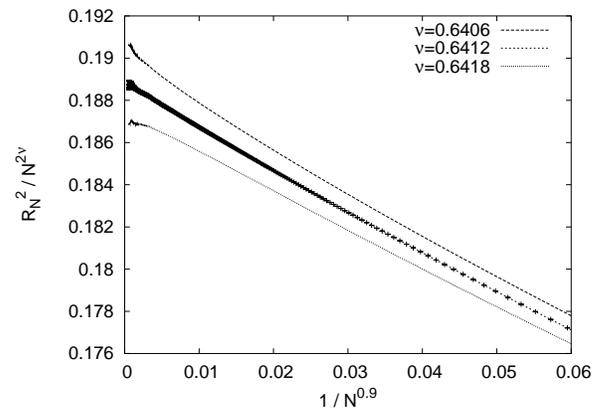,width=5.5cm,angle=270}
   \caption{Same data as in Fig.~\ref{rn2d}, but plotted against $1/N^{0.9}$.}
\label{rn2da}
\end{center}
\end{figure}

For these corrections one expects
\be
    Z_N \sim \mu^ N  N^{-\theta}(1+b_Z N^{-\Delta}+\cdots)   \label{ZNC}
\ee
and
\be
    R^2_N \sim N^{2\nu}(1+b_R N^{-\Delta}+\cdots) \;,     \label{RNC}
\ee
where $\Delta$ is the correction exponent~\cite{Adler}, $b_Z$ and $b_R$ 
are non-universal amplitudes, and the dots stand for higher order terms 
in $1/N$. Notice that $\Delta$ is universal, and is the same in both
equations. There are 
several methods discussed in the literature for estimating $\Delta$. We 
estimated it by plotting $\ln Z_N - aN + \ln N$ and $R^2_N /N^{2\nu}$ 
against $x\equiv 1/N^\delta$. Straight lines are expected near $x=0$ if
and only if $\delta=\Delta$. We could not find a value of $\delta$ 
where these lines were absolutely straight for all $x$, but the straightest
behaviour near $x\approx 0$ was obtained with $\delta\approx 0.9$, see 
Figs.~\ref{zn2da} and \ref{rn2da}.

 \begin{table*}
 \begin{center}
 \caption{Main results for site animals. For convenience we also give in the second column the critical $p$-values
     for site percolation.}
   \label{table1}
 \begin{ruledtabular}
 \begin{tabular}{ccllllll}
$d$ &  $p_c$   & {$a=\ln \mu$} & {$\theta$} & {$\nu$} & $\frac{\theta}{(d-2)\nu+1}$ & {$\Delta$} & Method\\
\hline
 2  & 0.5927  & 1.4018155(30)  & 1$^{a),c)}$   &  0.6412(5)  &   --   &  0.9(1)    & present work\\
    &         & 1.401815696(5) &            &  0.64115(5) &   --   &  1.0       & Series~\cite{Jensen03,Jensen00}\\
    &         &                &            &  0.642(10)  &   --   &  0.65(20)  & MC~\cite{You98}\\

\hline
 3  & 0.3116 & 2.1218588(25)  & 3/2$^{a),c)}$  & 1/2$^{a),c)}$  &  1$^{a),c)}$  &  0.75(8)      & present work (partially constr.)\\
    &        & 2.1218592(20)  & 3/2$^{a),c)}$  & 1/2$^{a),c)}$  &  1$^{a),c)}$  &  1$^{a),c)}$  & present work (constrained)\\
    &        & 2.120(2)       &             &             &            &            & Series~\cite{Guttmann78}\\
    &        &                & 1.502(3)    &             &            &  1$^{a),c)}$  & Series~ \cite{Fisher}   \\
    &        &                &             & 0.498(10)   &            &  0.54(12)  & MC~\cite{You98}\\
\hline
 4  & 0.1968 & 2.587858(6)    & 1.835(6)    & 0.4163(30)  &  1.001(7)  &  0.57(8)       & present work (unrestricted)\\
    &        & 2.5878583(40)  & 1.833(5)    & 0.4181(25)  &  0.998(4)  &  5/6$^{a),c)}$ & present work (partially constr.)\\
    &        & 2.5878583(30)  & 1.834(4)    & 0.417(2)    &  1$^{a),c)}$  &  5/6$^{a),c)}$ & present work (partially constr.)\\
    &        & 2.5878483(30)  & 11/6$^{a),c)}$ & 5/12$^{a),c)}$ &  1$^{a),c)}$  &  5/6$^{a),c)}$ & present work (constrained)\\
    &        &                & 1.839(8)    &             &            &  5/6$^{a),c)}$ & Series~\cite{Fisher}\\
    &        & 2.6012(15)     &             &             &            &             & MC~\cite{Lam} \\
    &        &                &             & 0.415(11)   &            &  0.46(11)   & MC~\cite{You98}\\
\hline
 5  & 0.1407 & 2.9223194(60)  & 2.080(7)    & 0.359(4)    &  1.001(9)  & 0.47(7)      & present work (unrestricted) \\
    &        & 2.9223205(30)  & 2.0815(60)  & 0.3605(20)  &  1$^{a),c)}$  & 0.622$^{b),c)}$ & present work (constrained)\\
    &        &                & 2.0877(25)  &             &            & 0.622(12)    & Series~\cite{Fisher}\\
    &        &                & 2.0807      &             &            &              & $\epsilon$-expansion~\cite{Fisher}\\
    &        &                & 2.10(3)     & 0.367(11)   &  1$^{a),c)}$  & 0.65(15)     & Series~\cite{Adler} \\
    &        & 2.899(9)       &             &             &            &              & expansion in $1/(2d-1)$ \cite{Peard95}\\
    &        & 2.940(15)      &             &             &            &              & MC~\cite{Lam}  \\
    &        &                &             & 0.359(11)   &            & 0.40(14)     & MC~\cite{You98}\\
\hline
 6  & 0.1090 & 3.1785245(40)  & 2.261(12)   & 0.315(4)    &  1.000(12) & 0.39(6)      & present work (unrestricted)\\
    &        & 3.178521(3)    & 2.256(8)    & 0.314(2)    &  1$^{a),c)}$ & 0.412$^{b),c)}$ & present work (constrained)\\
    &        &                & 2.2648(15)  &             &            & 0.412(8)     & Series~\cite{Fisher} \\
    &        &                & 2.2649      &             &            &              & $\epsilon$-expansion~\cite{Fisher}\\
    &        &                & 2.30(4)     & 0.325(10)   &  1$^{a),c)}$  & 0.5(2)       & Series~\cite{Adler} \\
    &        & 3.172(3)       &             &             &            &              & expansion in $1/(2d-1)$ \cite{Peard95}\\
    &        & 3.20(2)        &             &             &            &              & MC~\cite{Lam} \\
    &        &                &             & 0.321(19)   &            & 0.34(13)     & MC~\cite{You98} \\
\hline
 7  & 0.0889 & 3.384080(5)    & 2.40(2)     & 0.282(5)    & 0.996(20)  & 0.26(6)      & present work (unrestricted)\\
    &        & 3.384079(3)    & 2.390(9)    & 0.278(2)    & 1$^{a),c)}$  & 0.205$^{b),c)}$ & present work (constrained)\\
    &        &                & 2.402(5)    &             &            & 0.205(5)     & Series~\cite{Fisher} \\
    &        &                & 2.4999      &             &            &              & $\epsilon$-expansion~\cite{Fisher}\\
    &        &                & 2.41(3)     & 0.282(6)    &  1$^{a),c)}$  & 0.4(2)       & Series~\cite{Adler}\\
    &        & 3.382(1)       &             &             &            &              & expansion in $1/(2d-1)$ \cite{Peard95}\\
    &        & 3.41(1)        &             &             &            &              & MC~\cite{Lam}\\
    &        &                &             & 0.291(11)   &            & 0.35(7)      & MC~\cite{You98}\\
\hline
 8  & 0.0752 & 3.554827(4)    & 5/2$^{a),c)}$  & 1/4$^{a),c)}$  &  1$^{a),c)}$  & 0 (+logs)    & present work  \\
    &        & 3.5544(7)      &             &             &            &              & expansion in $1/(2d-1)$ \cite{Peard95}\\
\hline
 9  & 0.0652 & 3.700523(10)   & 5/2$^{a),c)}$  & 1/4$^{a),c)}$  &  1$^{a),c)}$  & 0.25(5)     & present work \\
 \end{tabular}
 \end{ruledtabular}
 \end{center}
 \flushleft{$^{a)}$ Exact value}\\
 $^{b)}$ From~\cite{Fisher} \\
 $^{c)}$ Used as constraint in the fit
\end{table*}

We thus conclude that $\Delta = 0.9\pm 0.1$ which suggests that the leading 
corrections to scaling are analytic ($\Delta = 1$ exactly).  This is in 
agreement with the exact enumerations of \cite{Jensen00,Jensen01,Jensen03}
and with the exactly known correction exponent for the Lee-Yang problem 
\cite{Adler,Fisher}, but disagrees with the Monte Carlo estimate 
$0.65\pm 0.20$ of \cite{You98}. Notice that originally \cite{Parisi} 
the connection between the Lee-Yang and animal problems was established 
only for the leading terms, and therefore the authors of \cite{Adler} 
suggested not to use the Lee-Yang correction to scaling exponents for 
animals. But the recent proof of \cite{Imbrie} gives an exact mapping 
between two models in the respective universality classes, and therefore 
we should use the mapping also for the corrections to scaling. 

The critical exponent $\nu$ and the growth constant $\mu$ can
be read off Figs.~\ref{zn2da},~\ref{rn2da}, and are reported in 
Table~\ref{table1}. The latter contains also our main results for all 
other dimensions. We see that our estimates for $\mu$ and $\nu$ are 
still much worse than the results obtained by the extremely long exact 
enumerations of Jensen, but they are more precise than all other previous
estimates.

We have also made ``unbiased" fits where we did not assume the theoretical
values $\theta=1$. We do not show details, we just mention that our data 
would seem to exclude $|\theta-1| > 0.002$.

\begin{figure}
  \begin{center}
   \psfig{file=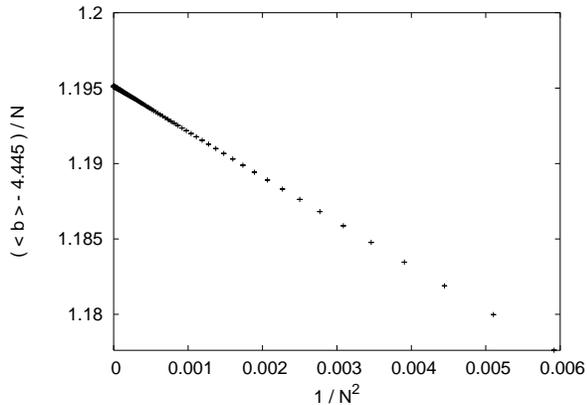,width=5.5cm,angle=270}
   \caption{Average number of boundary sites per cluster site for 2-$d$ animals,
    plotted against $1/N^2$. To reduce finite size effects we actually subtracted
    4.445 units from $\langle b \rangle$, before dividing it by $N$. Statistical 
    errors are much smaller than symbol sizes.}
\label{b2d}
\end{center}
\end{figure}

Finally, we show in Fig.~\ref{b2d} the average numbers of boundary sites. More precisely, 
with $\langle b \rangle$ being this average, we plot $(\langle b \rangle - 4.445)/N$ 
against $1/N^2$. Subtracting 4.445 units was done in order to reduce finite size 
effects. Without the very large correction $4.445/N$, this term would dominate
any other correction term, and would mask in particular any possible non-analytic 
correction. The linear shape of the curve suggests that there are no
non-analytic corrections, and that the next to leading term is $\sim 1/N^2$, but 
the data are too poor to allow a firm conclusion.

\subsection{$d>2$}

\begin{figure}
  \begin{center}
    \psfig{file=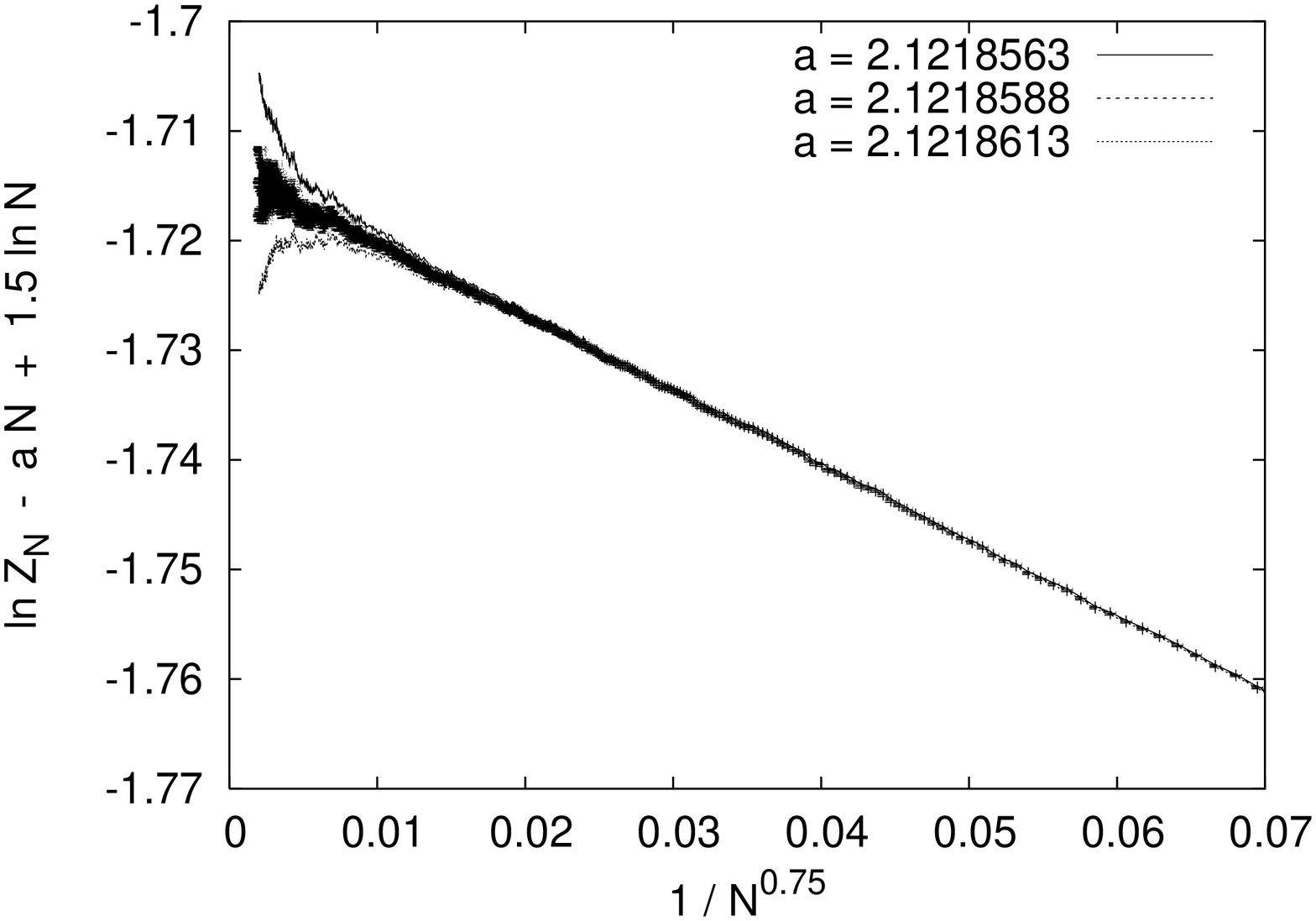,width=5.5cm,angle=270}
    \psfig{file=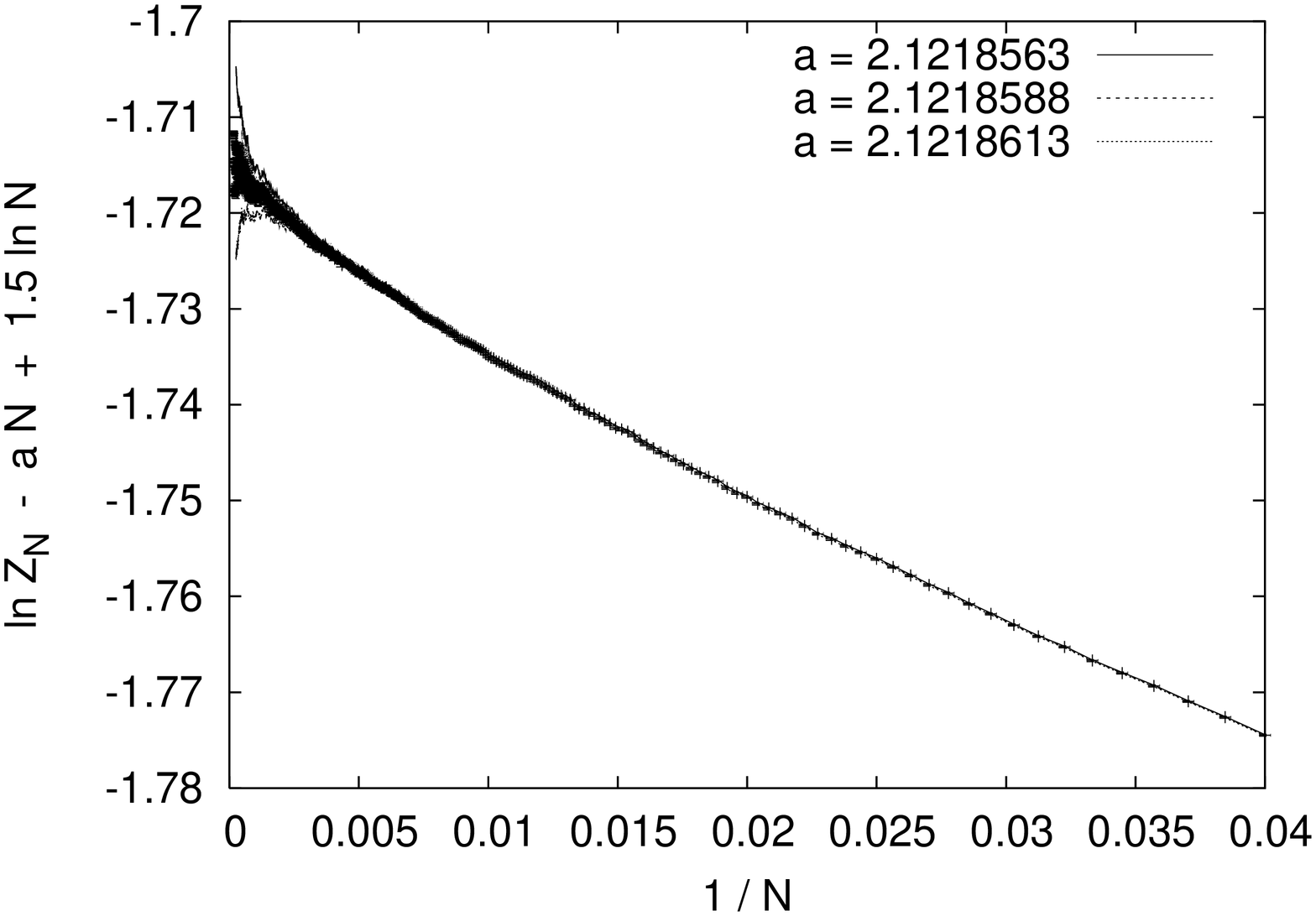,width=5.5cm,angle=270}
   \caption{Similar to Fig.~\ref{zn2da}, but for $d=3$. The straightest 
    curve was now obtained by plotting the data against $1/N^{0.75}$ 
    (upper panel). In the lower panel, the same data are plotted against
    $1/N$.}
\label{zn3da}
\end{center}
\end{figure}

\begin{figure}
  \begin{center}
   \psfig{file=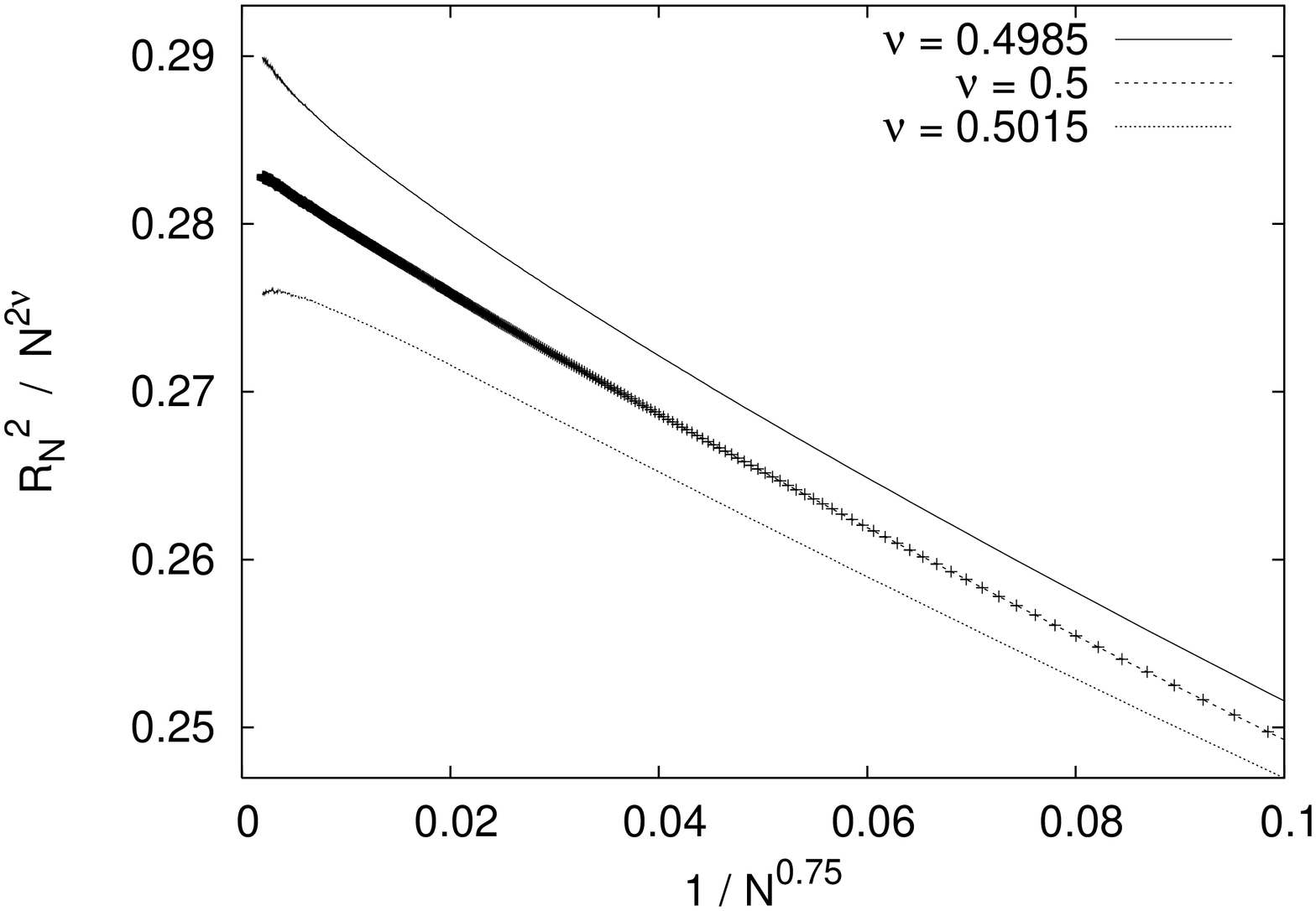,width=5.5cm,angle=270}
   \psfig{file=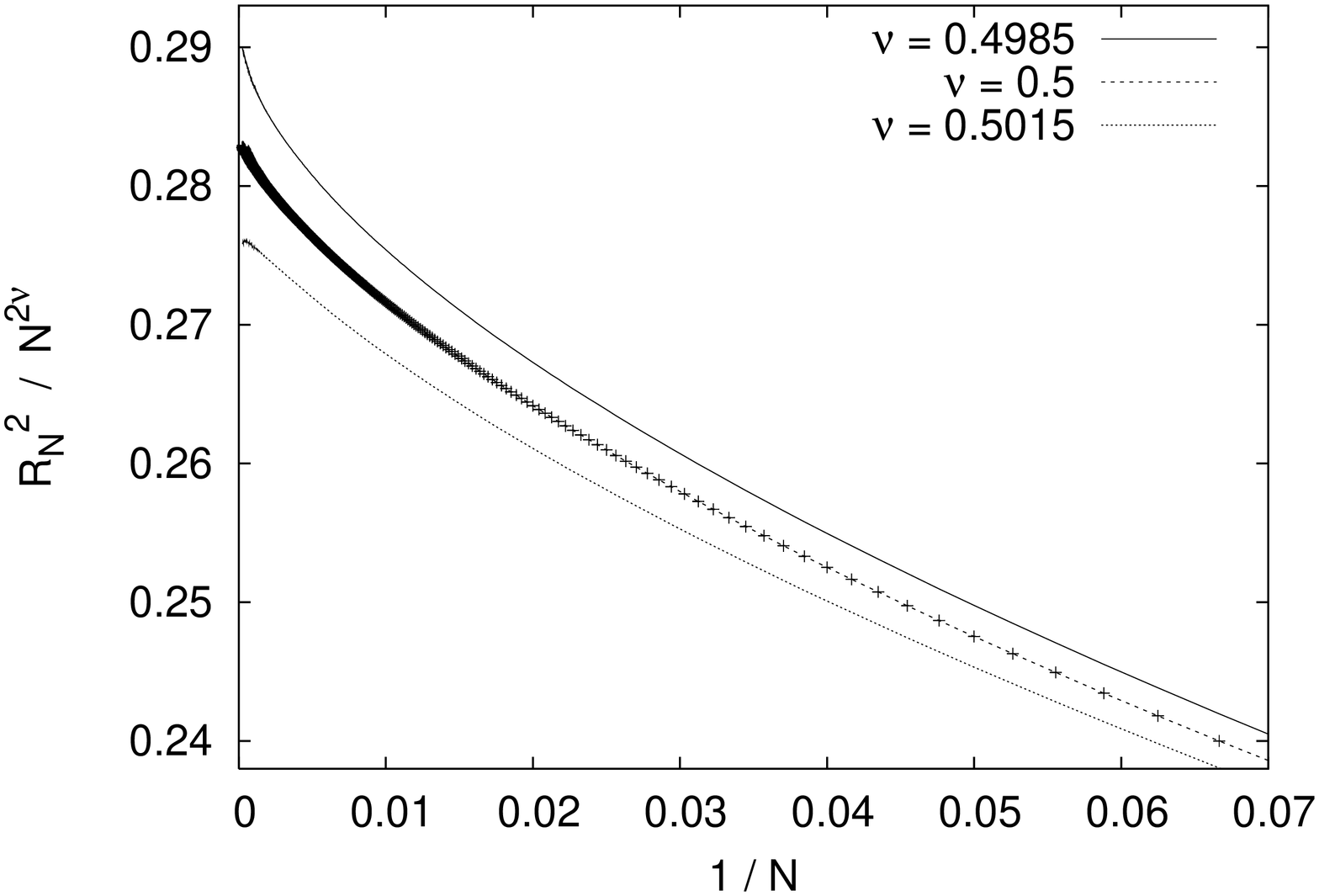,width=5.5cm,angle=270}
   \caption{Similar to Fig.~\ref{rn2da}, but for $d=3$. In the upper panel,
    $R^2_N/N^{2\nu}$ is plotted against $1/N^{0.75}$, in the lower panel it
    is plotted against $1/N$.}
\label{rn3da}
\end{center}
\end{figure}

For $d=3$ we show the data for $Z_N$ and $R_N^2$ in Figs.~\ref{zn3da} 
and \ref{rn3da}, plotted in the same way as in Figs.~\ref{zn2da} 
and \ref{rn2da}. Now the straightest line is clearly obtained 
for $\delta<1$, i.e. there definitely seem non-analytic corrections to 
scaling. The best fit was obtained with $\Delta=0.75\pm 0.08$ (upper panels
in Figs.~\ref{zn3da} and \ref{rn3da}). But from 
the Lee-Yang problem we know \cite{Fisher} that we have $\Delta=1$ also 
in $d=3$. As seen from the lower panels in Figs.~\ref{zn3da} and \ref{rn3da}, 
where these data 
are plotted against $1/N$, this is clearly not supported by our data. But we 
cannot, of course, exclude the possibility that this is due to very large 
higher order corrections to scaling. 
In view of this we show in Table~\ref{table1} two fits, one 
unrestricted where $\Delta$ is fitted from the present data and one constrained 
fit where $\Delta=1$ is imposed. In both fits the Parisi-Sourlas condition 
$\theta=1+(d-2)\nu$ and the exact values $\nu=1/2$ and $\theta=3/2$ were 
also used as constraints. Fits without imposing $\nu=1/2$ and $\theta=3/2$, 
and without assuming the Parisi-Sourlas relation, gave bigger errors for 
the growth constant, but gave exponents in full agreement with the predictions: 
$\nu=0.500\pm 0.002$ and $\theta=1.500\pm 0.001$. 

\begin{figure}
  \begin{center}
    \psfig{file=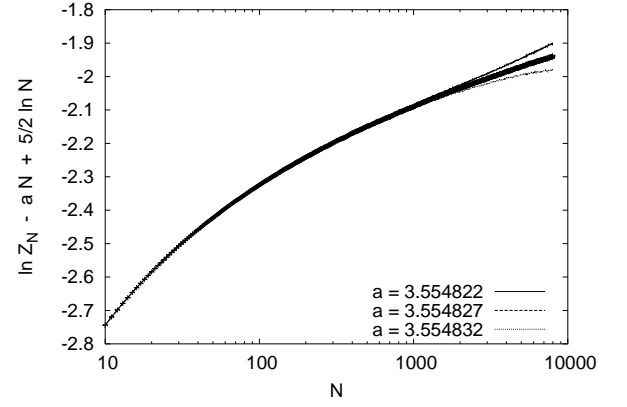,width=5.5cm,angle=270}
   \caption{$\ln Z_N-aN + 2.5 \ln N$ for $d=8$, plotted against $\ln N$,
    with three different values of $a$: 3.554822, 3.5548327, and 3.554832
    (top to bottom). Error bars are again plotted only for the central 
    curve.}
\label{fig8-d.zn}
\end{center}
\end{figure}

\begin{figure}
  \begin{center}
    \psfig{file=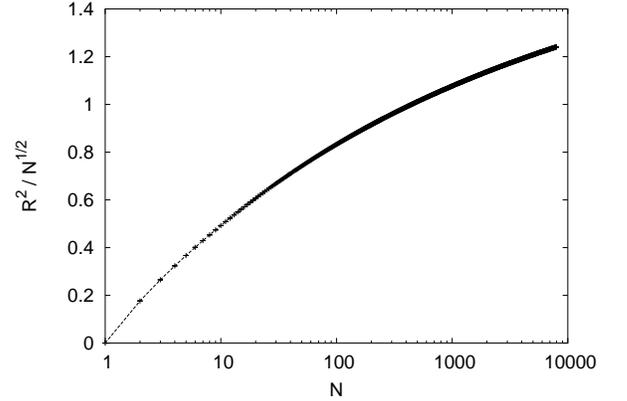,width=5.5cm,angle=270}
   \caption{Plot of $R_N^2/\sqrt{N}$ against $\ln N$ for animals in $d=8$.}
\label{fig8-d.rn}
\end{center}
\end{figure}

\begin{figure}
  \begin{center}
    \psfig{file=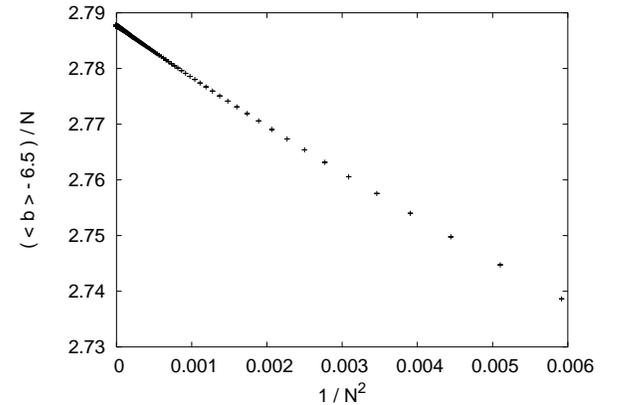,width=5.5cm,angle=270}
   \caption{$(\langle b\rangle - 6.5)/N$, where $\langle b\rangle$ is the 
    average number of boundary sites per cluster site for 3-$d$ animals,
    plotted against $1/N^2$. Statistical errors are much smaller than 
    symbol sizes.}
\label{b3da}
\end{center}
\end{figure}

\begin{figure}
  \begin{center}
   \psfig{file=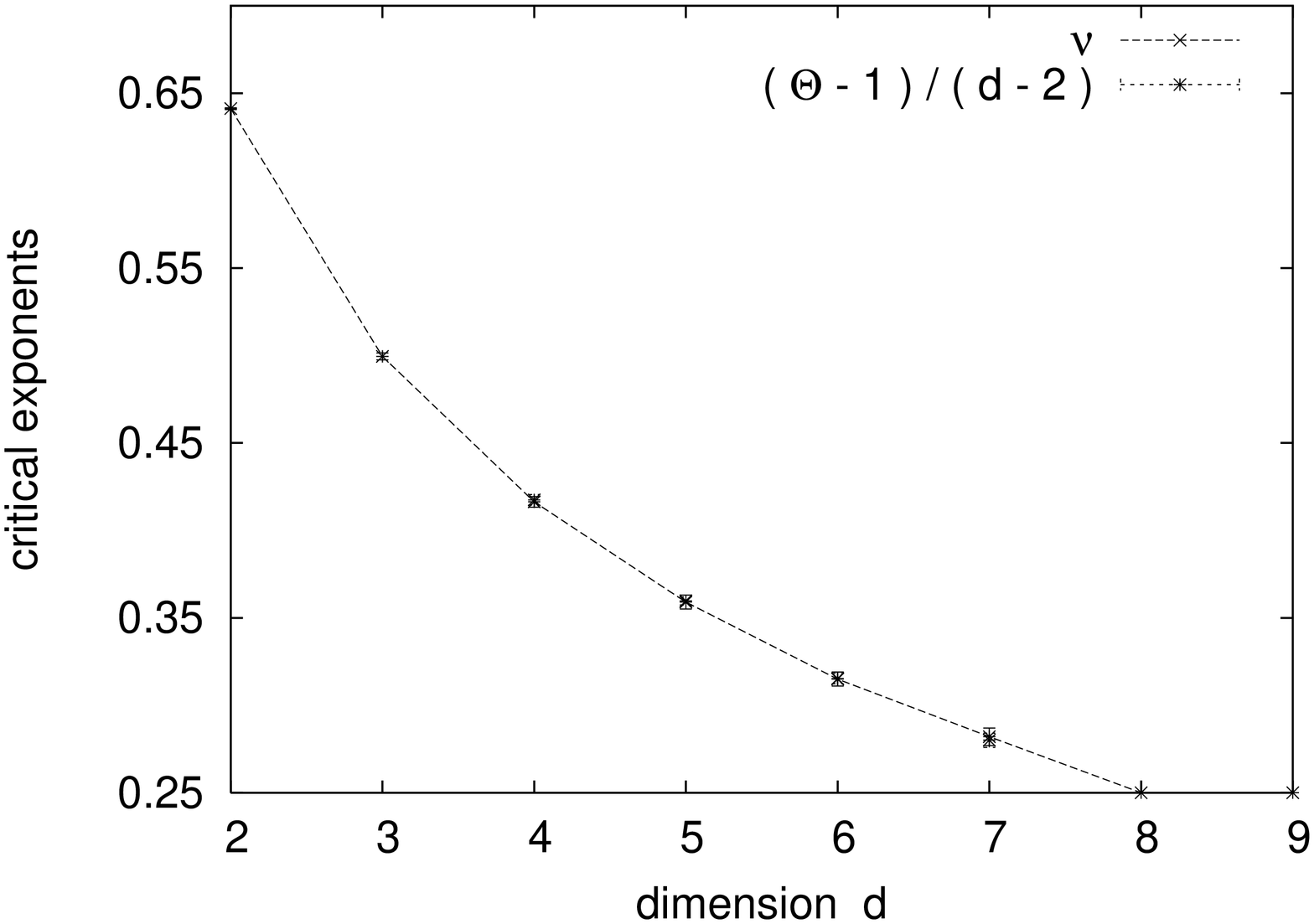,width=5.5cm,angle=270}
   \caption{The critical exponents $\nu$ and $(\theta-1)/(d-2)$ against $d$.}
\label{nutheta}
\end{center}
\end{figure}

The problem with the correction to scaling exponent is obviously due to 
large sub-leading corrections. It persists also in higher dimensions. For 
$d=4$, e.g., we estimated $\Delta=0.57\pm 0.08$, while the exact value
obtained from the Lee-Yang problem is $\Delta=5/6$. We present therefore 
in Table~\ref{table1} four different fits with various constraints: One 
completely unrestricted, another with $\Delta=5/6$ imposed, a third with the 
Parisi-Sourlas relation imposed in addition, and a final one with even the 
values $\theta=11/6$ and $\nu=5/12$ fixed. Notice that the growth constant 
can be obtained without knowing $\theta$, if the Parisi-Sourlas relation is 
assumed: In this case $N Z_N R_N^{d-2} \sim \mu^N(1+const/N^\Delta + \ldots)$.
From the values listed in Table~\ref{table1} we see that all four fits are 
mutually consistent.

Similar fits were also made for $d=5,6$ and 7, but we list in Table~\ref{table1}
only the results of the unrestricted and of the completely restricted fits.
In all cases the agreement between the fits is very good, showing the 
consistency of the data.

For $d=8$ we show in Figs.~\ref{fig8-d.zn} and 
\ref{fig8-d.rn} the data for free energies and for gyration radii, plotted 
against $\ln N$. From these plots one sees clearly that the data agree with 
the predicted exponents $\theta = 5/2$ and $\nu = 1/4$. But there are very large 
(presumably logarithmic) corrections, compatible with the fact that $d=8$ is 
the upper critical dimension. We have not tried to make a detailed fit to 
these corrections, since we are not aware of any theoretical prediction 
beyond the leading order \cite{Adler}, and because verifying logarithmic 
corrections is notoriously difficult.

 \begin{table}
 \begin{center}
 \caption{Asymptotic ratios between boundary and cluster sites,
     $\lim_{N\to\infty}\langle b \rangle / N$ (column \#2); ratios between 
     partition sums of bulk and wall grafted animals (column \#3); and 
     average number of contact of wall-grafted animals with that wall 
     (column \#4).
}
   \label{table2}
 \begin{ruledtabular}
 \begin{tabular}{c|ccc}
 dimension    & $\lim_{N\to\infty}\langle b \rangle / N$ & $\lim_{N\to\infty}Z_N'/Z_N$ &
          $\langle m \rangle$ \\ \hline
    2     &     1.1951(1) &  1.987(8) & 2.892(2) \\
    3     &     2.7877(1) &  2.97(3)  & 5.07(1) \\
    4     &     4.5859(2) &  3.98(5)  & 7.50(5) \\
    5     &     6.4909(2) &  4.91(4)  & 10.12(6) \\
    6     &     8.4503(1) &-- & -- \\
    7     &    10.4363(2) &-- & -- \\
    8     &    12.4346(1) &-- & -- \\
    9     &    14.4378(2) &-- & -- \\
 \end{tabular}
 \end{ruledtabular}
 \end{center}
 \end{table}

In Fig.~\ref{b3da}, the average number of boundary sites are plotted in a 
way similar to Fig.~\ref{b2d}. This time an even bigger term $6.5/N$ had 
to be subtracted, in order to see any possible non-analytic term. The fact 
that the curve is reasonably straight when plotted against $1/N^2$ suggests 
again (as for $d=2$) that there is no non-analytic correction term.
Our estimates for the critical exponents and for the growth constant are given 
in Table~\ref{table1}. 

The estimates for $\nu$ and for $(\theta-1)/(d-2)$ obtained by the unrestricted
fits are also shown in 
Fig.~\ref{nutheta}. According to Parisi and Sourlas, they should coincide.
The agreement is practically perfect. Moreover, both estimates have roughly
the same errors, and estimating $\nu$ indirectly, using Eq.~(\ref{theta}), 
seems to give slightly smaller errors for $d\geq 5$ than the direct 
measurement. For $d=6$ and $d=7$ our results are in very good agreement 
with the $\epsilon( = 8-d)$-expansion results of \cite{Alcantara}. Our 
results are also in very good agreement ($\leq 1\sigma$) with the series 
expansion of the {\it binary Gaussian molecule mixture} of \cite{Fisher}, 
which gives for high dimensions the most precise previous exponents for the 
Lee-Yang problem.

\begin{figure}
  \begin{center}
    \psfig{file=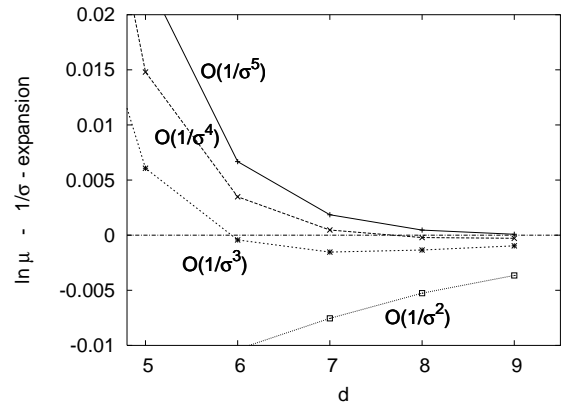,width=5.5cm,angle=270}
   \caption{Difference between the measured values of $\ln \mu(d)$ and successive 
    truncations of the expansion Eq.~(\ref{sigma}), plotted against $d$. Error 
    bars are smaller than the sizes of the symbols.}
\label{sigma-expansion}
\end{center}
\end{figure}

The growth constants seem to grow linearly with dimension, 
\be 
   \mu(d) \approx 5.49d - 8.94
\ee
for large $d$, although there are small but statistically significant deviations.
More precise expressions for the large-$d$ behaviour are obtained from an 
expansion of $\ln \mu$ in powers of $1/\sigma$ with $\sigma = 2d-1$ 
\cite{Peard95}:
\bea
   \ln \mu(d) & = & \ln \sigma +1 - {2\over \sigma}       \label{sigma} \\
  & -&\! {79\over 24\sigma^2} -
    {317\over 24\sigma^3} - {18321\over 320\sigma^4} -
    {123307\over 240\sigma^5} + O(\sigma^{-6}). \nonumber 
\eea
A comparison of our data with different truncations of this expansion is 
shown in Fig.~\ref{sigma-expansion}. This comparison suggests strongly that
the expansion is only asymptotic: For any fixed $d$, there is an order beyond 
which the expansion gives values smaller than the true value, and continues to 
decrease with increasing order.

Our estimates for the asymptotic number of boundary 
sites per cluster site are given in Table~\ref{table2}. The latter were
all obtained by assuming no non-analytic corrections to scaling, since 
our data can be fitted for all $d$ to $\langle b \rangle/N = \beta +\beta_1/N
+\beta_2/N^{\Delta'}$ with $\Delta'\approx 2$. For large $d$, our data 
seem to scale as 
\be
   \beta(d) = \lim_{N\to\infty}\langle b \rangle / N \sim 2d-const.
\ee

\section{Animals Attached to a Wall}

\subsection{Athermal Walls}

In this section we will consider $d$-dimensional animals grafted with 
one monomer to an impenetrable planar wall modeled by a hyperplane 
$x_d=0$. For this case, it was shown in \cite{DeBell} that the 
partition sum, written now $Z'_N$ instead of $Z_N$, scales as
\be
   Z'_N \sim \mu^N N^{-\theta'}
\ee
with the same $\mu$ as in the bulk, and with \cite{footnoteDeBell}
\be
   \theta' = \theta.
\ee

The last equation has a very simple heuristic explanation. Let us 
first map lattice points ${\bf x} = (x_1\ldots x_d)$ on a lattice 
of size $L^d$ onto integers 
\be
   i_{\bf x} = x_1 + x_2L \ldots x_dL^{d-1} 
\ee
(we actually used this in our codes to index points by a single 
integer, as this simplifies programming and makes memory access faster).
Consider now the problem of counting all animals restricted to
the half space $I_+ = \{{\bf x}: i_{\bf x} \geq i_0\equiv i_{{\bf x}_0}\}$
and positioned such that the site ${\bf x}_0$ belongs to the animal.
On the one hand, this means just that we consider animals with fixed
positions: for each shape we consider only that animal whose ``smallest" 
point is ${\bf x}_0$. Since we had counted only once all cluster shapes
related by translations, this means that the partition sum obtained 
now is exactly equal to $Z_N$.
On the other hand this model is equivalent to the animal being grafted 
to an impenetrable wall located at $x_d\approx x_{d,0}$ which is however 
not quite flat: $x_d$ jumps from $x_{d,0}+1$ to $x_{d,0}$ when any one 
of the other coordinates $x_j$ ($j=1,\ldots d-1$) goes through $x_{j,0}$. 
The proof of \cite{DeBell} then just shows that the scaling behaviour 
is independent of these steps, and is the same as for a flat surface.
In addition, this argument shows that $Z'_N>Z_N$ for all $N$ \cite{DeBell}.
Indeed, one easily sees that the ratio $\langle m_N\rangle_0 =Z'_N/Z_N$ is just the 
average number of contacts a {\it free} animal in the bulk would have with
a flat imaginary wall placed just below it. This is {\it not} equal
to the average number of contacts $\langle m_N\rangle$ 
of a {\it grafted} animal with its wall, because the latter animals 
are counted $m$ times if they have $m$ contacts: If we denote by $Z_{M,m}$ 
the number of configurations with $m$ sites in the bottom hyperplane and 
$N$ sites total (so that $Z_N = \sum_mZ_{M,m}$), and by $Z'_{M,m}$ the 
analogous quantity for grafted animals, then $Z'_{M,m} =m Z_{M,m}$.
Therefore, the average number of contacts of a grafted animal is given 
by the {\it second} moment of $m$ in the bulk ensemble divided
by the first moment, $\langle m_N\rangle \equiv \sum_m mZ'_{M,m}/\sum_mZ'_{M,m}=
\sum_m m^2 Z_{M,m}/\sum_m m Z_{M,m} \equiv \langle m^2_N\rangle_0/ \langle m_N\rangle_0$.

\begin{figure}
  \begin{center}
    \psfig{file=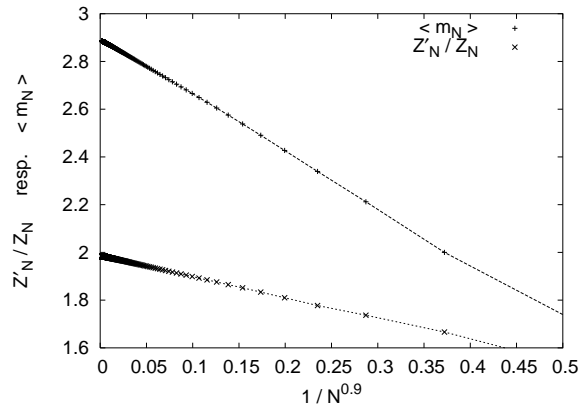,width=5.5cm,angle=270}
   \caption{Ratio $Z'_N/Z_N$ for 2-$d$ site animals, plotted against
    $1/N^{0.9}$ (lower curve) and average number of contacts of grafted 2-$d$
    animals (upper curve). Statistical errors are smaller than the sizes
    of the points.}
\label{zn-2d-ratio}
\end{center}
\end{figure}

In Fig.~\ref{zn-2d-ratio} we show the ratio $Z'_N/Z_N$ against 
$1/N^{0.9}$, together with the average number of contacts 
$\langle m_N\rangle$ of grafted animals, for $d=2$. 
We see again straight lines, showing that 
\be
   Z'_N/Z_N\;,\quad \langle m_N\rangle\; \sim\; const - {const\over N^{\Delta'}}
\ee
with $\Delta' = 0.9\pm 0.05$. Similar results were found for larger 
dimensions. The values of $\Delta'$ are close to those of $\Delta$
but somewhat larger, and we see no theoretical reason why they 
should be the same. We do not quote numbers since they are rather poorly
determined. The asymptotic values of $Z'_N/Z_N$ and 
$\langle m_N\rangle$ are given in Table~\ref{table2}. 

\subsection{Animals attached to an attractive surface}

The partition sum now is written as
\be
     Z_N^{(1)}(q) =\sum_{m=1}^N A_N(m) q^m   \label{ZNA}
\ee
where $A_N(m)$ is the number of configurations of lattice
animals with $N$ site having $m$ sites on the walls, and
$q=e^{\epsilon/KT}$ is the Boltzmann factor, $\epsilon>0$ is
the attractive energy between the monomer and the wall.

As $q \rightarrow 1$, there is no attraction between the
monomer and the wall, i.e. $Z_N^{(1)}(1) = Z_N'$. On the 
other hand it becomes clear that any cluster will collapse
onto the wall, if $q$ becomes sufficiently large. Therefore
we expect a phase transition from a grafted but otherwise 
detached to an adsorbed phase, similar to the transition 
observed also for linear polymers.

Exactly at the transition point $q=q_c$ we expect the usual
scaling laws
\be
   Z_N^{(1)}(q_c) \sim \mu^N N^{-\theta_s}    \label{Z}
\ee
and 
\be
   R_N \sim N^\nu .
\ee
In analogy to critical surface phenomena where this transition 
would correspond to the ``special" point \cite{Diehl}, we 
expect $\nu$ to be the same Flory exponent as in the bulk, 
while $\theta_s$ should be a new and independent exponent. 
The growth constant $\mu$, although being not universal, 
should be the same as in the bulk.

Away from the critical point we expect a scaling ansatz
\be
   Z_N^{(1)}(q) \sim \mu^N N^{-\theta_s} \Psi[(q-q_c)N^\phi],
\ee
with the crossover exponent $\phi$ being a second new exponent.
Taking the derivative of $\ln Z_N^{(1)}(q)$ with respect to 
$q$ and setting $q=q_c$ thereafter, we obtain for the average 
energy
\be
   E_N(q_c) = \langle \epsilon m \rangle \sim N^\phi.  \label{E}
\ee
Taking two derivatives we obtain for the specific heat per 
monomer near (but not exactly at) the critical point
\be
    C_N(q) = \frac{1}{NKT^2}(\langle (\epsilon m)^2\rangle -
               \langle \epsilon m \rangle^2) 
        \sim (q-q_c)^{-\alpha}   \label{C}
\ee
with 
\be
   \alpha=2-1/\phi ,
\ee
while
\be
   C_N(q_c) \sim N^{2\phi-1}.     \label{Cc}
\ee
In principle, all four scaling laws can be used to locate the 
critical value $q_c$. With conventional (Metropolis type) 
Monte Carlo simulations one cannot use easily Eq.~(\ref{Z}), 
since precise estimates of the partition sum are difficult to 
obtain. In this case it is usually Eq.~(\ref{C}) which is used.
With PERM we do have very precise estimates of $Z_N^{(1)}(q)$, 
and therefore we can use Eq.~(\ref{Z}), but we shall see that
it is indeed Eq.~(\ref{E}) which gives -- together with the 
two others -- the most precise estimate. This is very similar 
as for adsorption of linear polymers \cite{Hegger}.

In the following we shall assume $\epsilon=1$ without loss of 
generality. In order to compare with previous analyses we want to 
have specific heats for discrete values of $N$, but for a continuous
range of $q$. They are most easily obtained from histograms
\be
       P(m;q)=\sum_{\rm Config.}q^{m'} \delta_{m,m'}\; 
\ee             
which are normalized such that $Z_N^{(1)}(q)=\sum_m P(m;q)$.
Notice that we obtain from the simulations not only the shape
of the histogram, but also its absolute normalization, which makes 
it easy to combine two histograms obtained in runs with different 
nominal values of $q$. All we have to know are rough values of 
their relative statistical errors. These we can estimate from the 
number of tours which contribute to a particular value of $m$, 
$\Delta P(m;q) \propto 1/\sqrt{\# tours}$. Although this 
estimate is not very precise, it is fully sufficient to obtain 
smooth global histograms by joining histograms which cover
narrow regions in $m$.

\begin{figure}
  \begin{center}
   \psfig{file=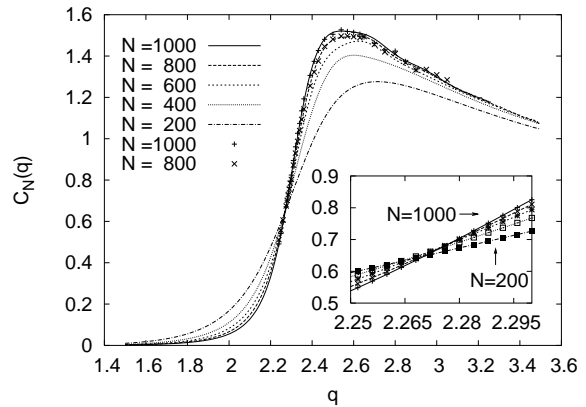,width=5.5cm,angle=270}
   \caption{Specific heats $C_N(q)$ per monomer for 2-$d$ site animals,
    plotted against $q$ for various values of $N$.
    Smooth curves show results obtained by histogram reweighing, points
    indicate results of single runs. The insert shows the region around 
    $q\approx 2.27$ where the curves intersect.}
\label{c2d}
\end{center}
\end{figure}

Specific heats for 2-$d$ animals with lengths up to $N=1200$ 
are shown in Fig.~\ref{c2d}. These data are very similar to the 
results of \cite{You01}, although the latter are for trees.
According to Janssen and Lyssy 
\cite{Janssen92,Janssen94,Janssen95} we expect $\alpha=0$, i.e. the 
specific heat curves for different $N$ should intersect exactly at 
the critical point. This gives roughly $q_c = 2.27$. But a close 
look at the insert in Fig.~\ref{c2d} reveals that these 
intersections slightly shift to larger $q$ as $N$ increases. 
Thus we have considerable corrections to scaling, preventing 
us from attributing error bars to this estimate.

\begin{figure}
  \begin{center}
    \psfig{file=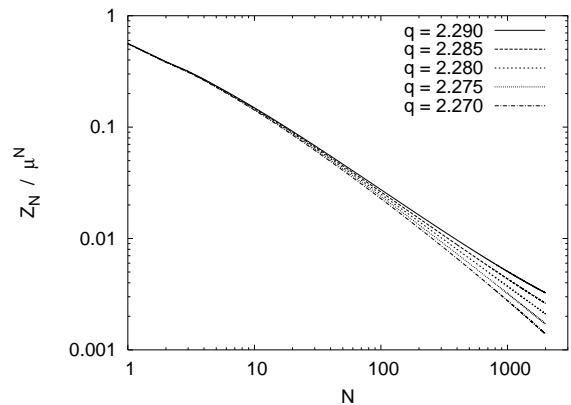,width=5.5cm,angle=270}
   \caption{$Z_N^{(1)}(q) / \mu^N$ against $N$ for various values of $q$, for
    site animals on the square lattice grafted to a wall.  Error bars here and
    in the following figures are comparable to the thickness of the lines.}
\label{z2d0}
\end{center}
\end{figure}

\begin{figure}
  \begin{center}
    \psfig{file=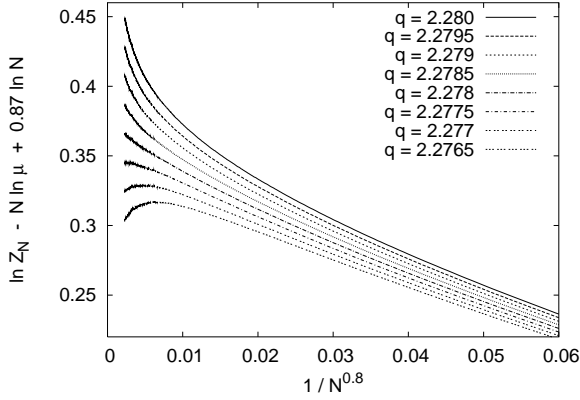,width=5.5cm,angle=270}
   \caption{Part of the data shown in Fig.~\ref{z2d0}, after interpolating
    them to a finer set of $q$-values, multiplying them with $N^{0.87}$, and
    taking their logarithm, plotted against $1/N^{0.8}$.}
\label{z2db}
\end{center}
\end{figure}

Alternatively, we turn towards the partition sum itself. In
Fig.~\ref{z2d0} we show log-log plots of $Z_N^{(1)}(q) / \mu^N$, for
various values of $q$ close to $q_c$. We see the expected power law, but
determining the critical point from this figure is difficult because of
the substantial corrections to scaling. We thus
multiply with an estimated power $N^{\theta_s}$ and plot the data against
$1/N^{0.8}$. The result is shown in Fig.~\ref{z2db}
(where we actually plot the logarithm on the $y$-axis). Notice that we did
not have to make a new estimate of $\mu$; rather, we could take the old and
very precise estimate. The value $\theta_s = 0.87$ was chosen so as to
give the best straight line for a suitably chosen $q_c$. Indeed,
from this plot we would conclude that $q_c \approx 2.278$.


\begin{figure}
  \begin{center}
    \psfig{file=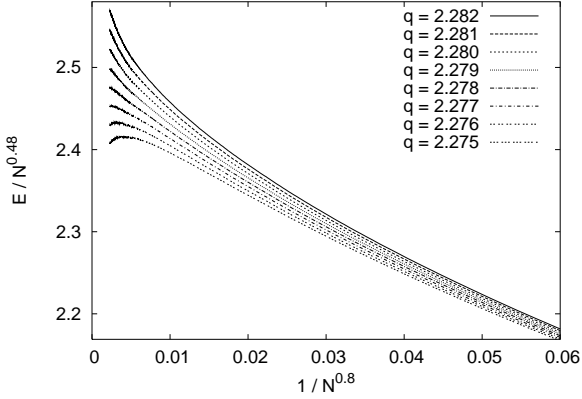,width=5.5cm,angle=270}
   \caption{$E/N^{0.48}$ against $1/N^{0.8}$ for various values of $q$, for
    site animals on the square lattice grafted to a wall.}
 \label{e2db}
\end{center}
\end{figure}

We now turn towards Eq.~(\ref{E}). In Fig.~\ref{e2db} we plotted
$E/N^{0.48}$ instead of $E/\sqrt{N}$. This was chosen because it suggests
that $ q_c \approx 2.278$, in agreement with the value obtained
above from $Z_N$. Assuming $\phi=1/2$, in contrast, would have given
$q_c \approx 2.283$ which would be incompatible with the data for $Z_N$.

\begin{figure}
  \begin{center}
    \psfig{file=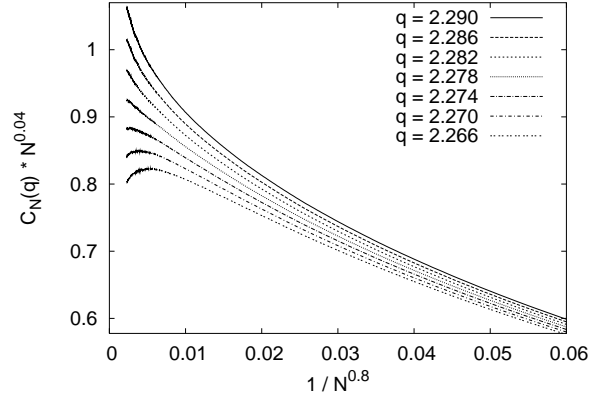,width=5.5cm,angle=270}
   \caption{Part of the data shown in Fig.~\ref{c2d}, but multiplied by $N^{1-2\phi}$
    and plotted as curves with fixed $q$ against $1/N^{0.8}$.}
\label{c2da}
\end{center}
\end{figure}

Finally, we plot $N^{1-2\phi} C_N(q)$
against $1/N^{0.8}$, in order to compare with Eq.(\ref{Cc}). As seen 
from Fig.~\ref{c2da} this is fully consistent with $q_c \approx 2.278$. 
If we had taken $\phi=1/2$, we would again get a too large estimate 
$q_c \approx 2.285$.

Summarizing, we obtain as our best estimates:
\be
   q_c = 2.2778\pm 0.0008\;,\quad \phi=0.480\pm 0.004
\ee
together with
\be
   \theta_s = 0.870 \pm 0.009\;,\quad \Delta_s = 0.8\pm 0.2.
\ee
The large error of $\Delta_s$ reflects the fact that the best estimates
obtained from the different observables would be quite different, suggesting
again large non-leading corrections. But this seems to have little effect
on the estimates of the other quantities. Since we believe that we have 
taken into account all systematic errors, we claim that the Janssen-Lyssy
conjecture $\phi=1/2$ is slightly but significantly violated in $d=2$.
The previous estimates $\phi=0.505(15)$ \cite{Queiroz} and $\phi=0.503(3)$
\cite{Vujic} most likely suffer from such systematic errors. On the other 
hand, our result is in agreement with the MC estimate $\phi=0.50(3)$ of
\cite{You01}. Surprisingly, our estimate for $\phi$ agrees within the error
bars with the most recent estimate of the cross-over exponent for unbranched
polymers attached to an attractive wall in 3 dimensions, 
$\phi_{\rm SAW} = 0.484(2)$ \cite{Grassberger04}, while $\phi_{\rm SAW} = 0.5$
for unbranched 2-d SAWs \cite{Burkhardt89}.

Before leaving this problem, we should discuss the gyration radii. Their
behaviour near the critical adsorption point is somewhat more complicated.
For $q>q_c$ the gyration radius scales as $R_N \sim N$. At $q\leq q_c$ we
expect it to scale as $R_N \sim N^\nu$ with the same value of $\nu$ as in
the bulk, as in other surface critical phenomena \cite{Diehl}. The effect
of the wall is only seen then in the amplitude $A = \lim_{N\to\infty} \langle R_N^2
\rangle /N^{2\nu}$. For $q<q_c$ it should be larger than for free animals, and for
$q=q_c$ we should expect it to be even larger, $A_{\rm bulk} < A_{\rm wall}
< A_c$. The reason is that the main effect of the wall is to squeeze the
animal in the direction perpendicular to the wall, which by the excluded
volume effect makes it more extended in the direction parallel to the wall.

This is a bit analogous to the case of an unbranched polymer between
two athermal walls \cite{vliet}:
If the distance $D$ between the walls is decreased, at first the shrinking of
the perpendicular extension dominates any increase parallel to the plates.
However, if $D$ is much smaller than the Flory radius, the stretching
parallel to the walls dominates, and $R_N$ increases in comparison to a free
polymer \cite{vliet}.

\begin{figure}
  \begin{center}
    \psfig{file=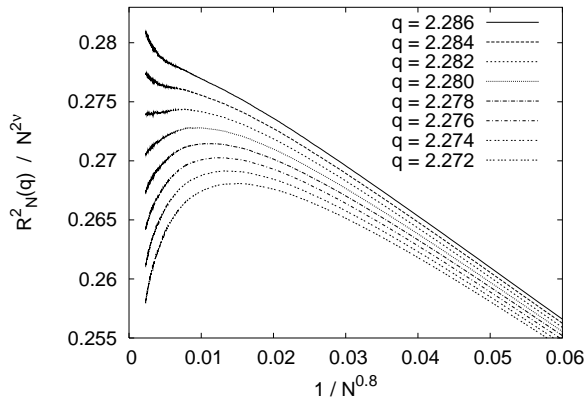,width=5.5cm,angle=270}
   \caption{Ratios $\langle R_N^2 \rangle / N^{2\nu}$, where $R_N$ is the 
    gyration radius of grafted 2-d animals, plotted against $1/N^{0.8}$.}
\label{R2d}
\end{center}
\end{figure}

Our data (Fig.\ref{R2d}) indicate that $R_N/N^\nu$ is larger than for animals
in the bulk (where $A_{\rm bulk} \approx 0.189$), and that it increases
with $q$. But at $q\approx q_c$ it is not monotonic in $N$: It increases with
$N$ until $N\approx 300$, and then decreases sharply. This strange behaviour
might have been expected from the analogy with unbranched self avoiding walks
between two athermal walls. For small $N$ and $q=q_c$ the effect of the
wall is strong, and the increase of the size parallel to the wall dominates.
But for $N\to \infty$ the effect of the wall becomes increasingly weaker,
and the stretching along the wall becomes less important. We verified that
it is indeed the slower increase of the parallel component which lets 
$R_N/N^\nu$ decrease for large $N$, but we found no similar effect in
simulations (unpublished) of unbranched polymers at the critical adsorption
point. Thus we have at the moment no good explanation for this effect.

 \begin{table}
 \begin{center}
 \caption{Critical Boltzmann factors, crossover exponents, and critical 
     exponents $\theta_s$ at the adsorption transition for site animals
     on simple (hyper-)cubic lattices grafted to a flat attractive wall.}
   \label{table3}
 \begin{ruledtabular}
 \begin{tabular}{c|ccc}
 dimension    & $q_c$ & $\phi$ & $\theta_s$ \\ \hline
    2     &  2.2778(8)   &    0.480(4) &  0.870(9) \\
    3     &  1.4747(6)   &     0.50(1) &  1.476(7)  \\
    4     &  1.2674(6)   &     0.50(2) &  1.91(1)   \\
    5     &  1.1786(5)   &     0.51(3) &  2.18(4)    \\
 \end{tabular}
 \end{ruledtabular}
 \end{center}
 \end{table}

For higher dimensions, the same kind of analysis as in $d=2$ gave the estimates 
given in Table~\ref{table3}. The main problem in these analyses is again that the 
best estimate for the leading correction to scaling exponents for the different 
observables did not quite agree with each other. This hints at the presence 
of more than one important
term in the scaling corrections, and it dominates the error estimates. The 
most remarkable result seen in Table~\ref{table3} is the perfect agreement with 
the Janssen-Lyssy prediction $\phi=1/2$ in all dimensions $\geq 3$. In particular,
it seems that the strong violation seen in \cite{Lam-Binder} for $d=3$ was due 
to an underestimation of finite size effects. Actually, Janssen and Lyssy 
had derived $\phi=1/2$ only for $d=3, 4$, and $\geq 8$, but not for $d=2$.

\begin{figure}
  \begin{center}
   \psfig{file=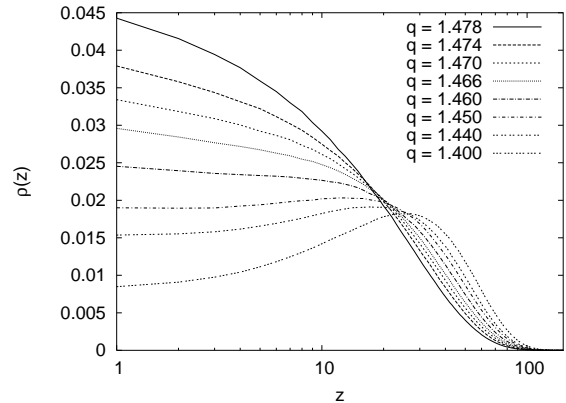,width=5.5cm,angle=270}
   \caption{Monomer density profiles $\rho(z)$ (with $z$ being the distance from 
    the wall) for a 3-$d$ animal grafted to an attracting wall.}
\label{e3dc}
\end{center}
\end{figure}

Finally, we show in Fig.~\ref{e3dc} the monomer density profile $\rho(z)$ for 
3-$d$ animals, where $z$ is the distance from the wall. Animal sizes for this 
figure were $N=4000$. All curves must of course go to zero for $z\to\infty$.
At temperatures far above $T_c$, i.e. at $q\ll q_c$, the entropic repulsion 
from the wall dominates and $\rho(z)$ has a maximum at a finite value of $z$.
In the adsorbed phase $\rho(z)$has its maximum at $z=0$ and decreases 
monotonically with $z$. Notice that the transition from non-monotonic to 
monotonic behaviour does not happen exactly at $q_c$, but for $q$ slightly smaller 
than $q_c$. Presumably this is a finite size effect, and the transition would 
happen at $q_c$ for much larger animals.

\section{Trees and Bond Animals}

\subsection{Site Trees}

The simplest modification of the codes presented so far is needed for 
simulating site trees. As we pointed out in the introduction, site trees
are site animals without loops. Thus the number of nearest neighbour pairs
is just $N-1$ for a tree of $N$ sites. It is easy to count the number 
of occupied nearest neighbour pairs as the cluster grows. We have
just to prune the growth as soon as this number is equal or larger than $N$.
Apart from that, pruning and branching is done exactly as before, and all
comments made in Sec.~2 about the efficiency of the algorithm apply also
to site trees. 

We made simulations only for 2-$d$ site trees, and only with rather modest 
statistics. Our results were fully in agreement with those of Jensen
\cite{Jensen01}. In particular we obtained $\mu=3.79527(4)$ after three
days of CPU time on a 3GHz Pentium, to be compared to the estimate
3.795254(8) obtained in \cite{Jensen01}.

\subsection{Bond Animals and Bond Trees}

In order to simulate bond animals and bond trees, one has to grow bond
instead of site percolation clusters. Cluster growth algorithms for bond 
percolation are very similar to those for site percolation and about as 
easy. One just has to remember that in bond percolation one often does 
not distinguish between clusters with the same configurations of sites, 
but with different bond configurations. However, for animals it is 
essential to make this distinction.

Let us denote by $k$ the number of non-bonded nearest neighbour pairs 
(often called `contacts' in this context), by $b$ the number of 
surface bonds, and by $m$ the number of established bonds between 
nearest neighbours. In the bond percolation ensemble, a cluster 
with these `quantum numbers' has a weight (cf. Eq.(~\ref{Pperc}))
\be
   P_{Nkbm} = p^m(1-p)^{b+k}\;.                         \label{wbp}
\ee
This is slightly more complicated than in the site percolation case, but 
one can follow the same strategy when using this to simulate (bond) animals.
We just have to replace the number of perimeter sites in the weight factor 
by $b+k$, and if we want to simulate trees, we have of course to prune all 
configurations which are not tree-like. Growth sites have to be replaced by
growth bonds. Moreover, it is a bit more natural to consider ensembles with 
fixed $m$, i.e. with fixed number of established bonds, rather than sites
\cite{Rensburg97}.

The heuristics worked out in Sec.~2 remain valid: one obtains much better
results when the trees are grown breadth first instead of depth first; 
one should use a fitness function $f({\cal C}) = (1-p)^{-g} W({\cal C})$, 
where $g$ is now the number of growth {\it bonds}; and one should 
simulate at a slightly subcritical value of $p$ which approaches $p_c$ 
as the trees to be simulated become larger and larger. For the same values 
of $N$, the optimal values of $p/p_c$ were however somewhat smaller. For 
$d=2$ and $N=1000$, e.g., best results were obtained with $p\approx 0.45$
(with $p_c=1/2$).

\begin{figure}
  \begin{center}
   \psfig{file=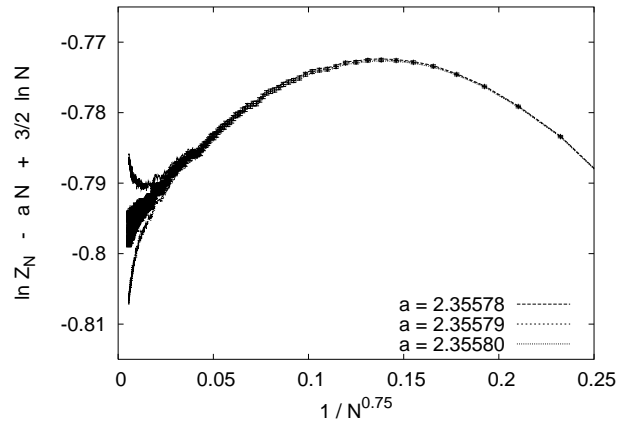,width=5.8cm,angle=270}
   \caption{Plot of $\ln Z_N - a N + 3/2 \ln N$ against $1/N^{0.75}$, for bond
    trees on the simple cubic lattice.}
\label{3d-bondtree-zn}
\end{center}
\end{figure}

Let us first discuss trees. In two dimensions, we obtained 
$\mu = 5.14276\pm 0.00002$. This is compatible with the best previous estimate, 
$\mu = 5.14339\pm 0.00072$ \cite{Rensburg03}, but more than an order of 
magnitude more precise. Our estimate is based on rather small clusters
($N_{\rm max}=500$), but very high statistics: The error of $\ln Z_N$ is 
$\Delta \ln Z_{N=500} = 0.000451$.
These simulations confirmed also that the leading correction to scaling 
exponent is between 0.9 and 1.0. In three dimensions we obtained $\mu = 
10.54646 \pm 0.00010$. This case is a bit special, since it has strange 
looking corrections to scaling, as seen from Fig.~\ref{3d-bondtree-zn}.
Apart from Fig.~\ref{R2d}, this is our clearest and most striking 
example showing that the corrections 
cannot be described by a single power, and illustrates the pitfalls in 
estimating correction to scaling exponents from poor data. In other cases, 
the existence of more than one power in corrections to scaling have often 
to be inferred less directly, e.g. by comparing different observables (see
Sec.~4B) or by invoking universality \cite{backbone}.

General animals were simulated only with somewhat lower statistics, since 
we are not aware of any other recent high statistics simulations.  On the 
square lattice we obtained $\mu = 5.20789\pm 0.00004$, to be compared with 
the previous series expansions estimate $ 5.208\pm 0.004$ of \cite{Guttmann78}. 
On the simple cubic lattice we got $\mu = 10.61539\pm 0.00006$, to be compared 
with $10.63\pm 0.05$ \cite{Alcantara} and $10.62\pm 0.08$ \cite{Gaunt78}. 
Mean-square gyration radii were e.g. $R_{N=499}^2 = 347.974 \pm 0.036$
in $d=2$ and $R_{N=999}^2 = 165.669 \pm 0.022$ in $d=3$, to be compared to 
the best previous MC estimates, $348.32 \pm 0.88$ and $166.03\pm 0.34$ 
\cite{Rensburg97} (the quantity $n$ displayed in the first lines of Tables A1 to 
A7 of \cite{Rensburg97} is not the number of bonds, as stated there, but one 
plus this number).

Finally we point out that we can also use bond percolation as a starting point 
for the simulation of site animals. We just have to use the fact that site 
animals are isomorphic to the subset of bond animals with maximal number of 
bonds for a given configuration of occupied sites. Using this we obtained for 
$d=2$ results in agreement with those of Sec.~3, but the algorithm was somewhat
less efficient than that based on site percolation.

\section{Animal Collapse}

\subsection{Collapse of Site Animals}

In order to describe collapsing animals and/or trees, one has to 
introduce attractive monomer-monomer interactions. Historically the 
first model of this type \cite{Dickman,Derrida-H} starts from site 
animals and introduces a contact energy for each `contact', where 
a contact is a pair of occupied nearest neighbour sites (notice that 
the definition of contacts used here differs slightly from that used for 
bond animals). Let us denote this energy as $-\epsilon$, and the 
corresponding Boltzmann factor as $q=\exp(\beta\epsilon) >1$. Let us 
furthermore denote by $m$ the number of contacts. The 
partition sum is then written as 
\be
   Z_N(q) = \sum_m C_{N,m} q^m
\ee
where $C_{N,m}$ is the number of different clusters with $N$ sites and $m$
contacts. If the clusters are embedded in some solvent, repulsive 
monomer-solvent interactions need not be included explicitly, since the 
number $s$ of solvent contacts satisfies ${\cal N}N = s+2m$, where ${\cal N}$
is the coordination number of the lattice (${\cal N} = 2d$ on a simple 
hypercubic lattice).

Simulations with PERM are straight forward. We just have to modify the 
weight factors $W_N$ to $W_{N,m} = W_N e^{\beta m}$. We found that again,
as for the previous case $q=1$, it was better to simulate clusters breadth
first than depth first. We also verified that it was advantageous to 
include a factor $\approx (1-p)^{-g}$ in the fitness functions, just as 
for athermal clusters. But the results were rather disappointing, at least
for low dimensions. This might seem at first surprising, given the fact 
that PERM works extremely well at the collapse transition of linear 
polymers in 3 dimensions \cite{g97}. But it is easy to see the reason 
for this difference. For linear polymers, $d=3$ is the upper critical 
dimension, and $\theta$ polymers form essentially random walks with very 
small logarithmic corrections. Thus starting off with random walks, PERM
can do with very few resampling steps. There is only one pruning or 
cloning needed for every 2000 simple forward steps \cite{g97}. Collapsing
site trees, at least in low dimensions, are however very different from 
site percolation clusters. A convenient observable to see this difference
is the average number $\langle m \rangle$ of contacts. For both models,
$\langle m \rangle$ is roughly proportional to $N$. But for site 
percolation on the simple cubic lattice one finds $\langle m \rangle /N 
\approx 0.15$, while the same number for collapsing animals (at 
$q \approx 3.22$ \cite{Lam88}) is $\approx 0.40$. Thus there is still a huge
amount of re-sampling needed, even more than for athermal animals 
where $\langle m \rangle/N \approx 0.065$.

\begin{figure}
  \begin{center}
   \psfig{file=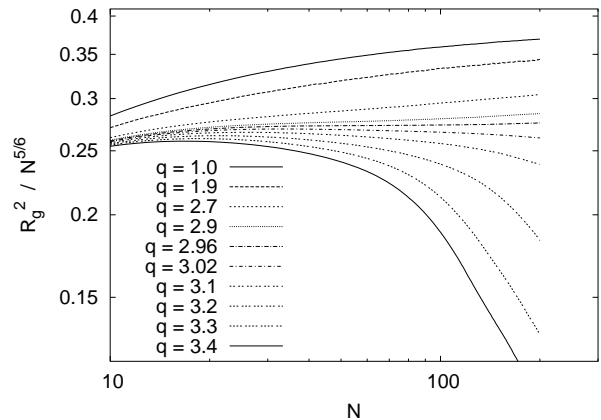,width=5.8cm,angle=270}
   \caption{Squared gyration radii of collapsing site animals in four
    dimensions divided by $N^{2\nu}$, where $\nu=5/12$ is the exact Flory
    exponent for non-collapsing animals.}
\label{fig_4d-collapse}
\end{center}
\end{figure}

For this reason we do not even show any data for the collapse in $d=2$ 
and $d=3$. The 
situation improves when $d$ is increased. Results for $d=4$ are shown 
in Fig. \ref{fig_4d-collapse}. We see a rather sharp collapse transition 
at $q=q_c = 2.98\pm 0.02$. It is hard to give precise estimates of any critical 
exponents from these data. But the Flory exponent $\nu$ seems to be the 
same as for athermal animals, within rather small errors. This would not 
be easy to understand, if it were exactly true. Anyhow, we are not 
aware of any theoretical prediction to compare this with. We are not even 
aware of any prediction of the upper critical dimension for this collapse
transition.

\subsection{Collapse of Bond Animals and Bond Trees}

Let us now switch attention to collapse models based on bond animals.
We have now two different possible 
interactions. Instead of the single parameter $\epsilon$ for the 
interaction strength in site animals, we can now introduce different
parameters $\epsilon_1$ for bonded and $\epsilon_2$ for non-bonded 
neighbour contacts, and define $y=e^{\epsilon_1}\;, \tau=e^{\epsilon_2}$. 
Notice that as before we do {\it not} have to introduce also an 
additional interaction with the solvent, because the number $b$ of 
surface bonds is not independent of $m,k$, and $N$: 
\be
   2m+2k+b={\cal N}N.      \label{coord}
\ee
We then define
\be
   Z_N(y,\tau) = \sum_{m,k} C_{N,m,k} y^m \tau^k \;.
\ee

The model discussed in the last subsection is obtained by taking the 
limit $\tau=0$ \cite{Henkel96}. In this limit only bond configurations 
with the maximal number of bonds (for a given configuration of sites) 
contribute to the partition sum. On the other hand, bond percolation 
without reweighing corresponds, due to Eqs.~(\ref{wbp}) and (\ref{coord}),
to the curve
\be
   y=p/(1-p)^2\;,\quad \tau=1/(1-p)\;,\qquad 0\leq p \leq 1 \;.
\ee
or, explicitly, 
\be
   y=\tau(\tau-1)\;,\qquad \tau\geq 1\;.
\ee

Critical bond percolation ($p_c=1/2$) corresponds to $y=\tau=2$.
Simulations should be very easy in the neighbourhood of this point, 
but they also should be not too difficult in the neighbourhood of the 
entire bond percolation line. The reason is simply that along the 
entire line one should not need much resampling. This should be enough
to obtain precise estimates for large parts of the phase diagram, 
and in particular to clarify the existence of two different collapsed
phases. Results will be given in a forthcoming paper.

\section{Conclusion}

We have shown that the basic idea of PERM, namely the recursive 
implementation of biased sequential sampling with re-sampling, can 
be applied also to lattice animals and lattice trees. These are two 
models for randomly branched polymers. Our algorithm is extremely 
efficient (except for the collapse of site animals), obviously much 
more efficient than previous Monte Carlo
algorithms. We applied it to simple (hyper-)cubic lattices in up to 
nine dimensions, but made also less complete simulations on bcc and 
fcc lattices. Our algorithm works indeed better in higher dimensions,
nevertheless we obtained high statistics results also for large 
animals in two dimensions. 

We verified a number of theoretical predictions. In particular, we verified 
the Parisi-Sourlas connection between entropic and Flory exponents, 
and we verified the values of these exponents whenever they are 
exactly known. We also verified that the cross-over exponents for 
branched polymer adsorption on plane walls is super-universal 
for $d>2$, as predicted some time ago (but not for $d=2$!), and we 
gave precise estimates of the 
other critical exponents at this adsorption transition.

There are a number of problems we did not yet study, although
our algorithm seems ideally suited for them, and which we plan 
to address in forthcoming papers. One is that of collapsing animals
where we hope to be able to verify or disprove the existence of 
two different collapsed phases. Another is the dependence on the 
wedge angle, of the entropic exponent of a 2-$d$ animal grafted at the 
tip of this wedge. In conformally invariant 2-$d$ theories this 
angle dependence can be predicted, but lattice animals are not 
conformally invariant.

Apart from these specific problems we believe that the present 
simulations have demonstrated again the power of sophisticated 
sequential sampling methods, and of PERM in particular. Although
there are certainly many problems where other MC strategies are 
more efficient, there are by now many examples where PERM seems
unchallenged by any other known method. Unfortunately (or, rather, 
fortunately for the livelyhood of the subject) it is hard to 
predict when PERM or a similar strategy will be the method of 
choice. But we are confident that lattice animals will not be the 
last such problem.

Acknowledgements: We thank Michael Fisher for bringing Ref.
\cite{Fisher} to our attention, and him and Buks van Rensburg
for very useful correspondence.


\begin{thebibliography}{99}
\bibitem{footnote0} The only previous exception was in ground state predictions
   for the HP protein model, where it was found useful to include a bias
   against H-P contacts in the population control \cite{HP-PRE,nicsympos}.
\bibitem{footnote1} Notice the very large values of $\ln W$. Much effort
   was spent to write the codes such that overflows were avoided.
\end{thebibliography}
\end{document}